\documentclass[twocolumn]{aastex63}

\usepackage{bm}
\usepackage{textcomp}
\usepackage{gensymb}
\usepackage{multirow}
\usepackage{wrapfig}
\usepackage{amsmath}
\usepackage{xcolor}
\usepackage{hyperref}

\def\iso#1#2{\mbox{${}^{#2}{\rm #1}$}}
\def\fe6#1{\iso{Fe}{6#1}}
\def\mn5#1{\iso{Mn}{5#1}}
\def\al2#1{\iso{Al}{2#1}}
\def\pu24#1{\iso{Pu}{24#1}}

\def\pfrac#1#2{\left( \frac{#1}{#2} \right)}

\submitjournal{The Astrophysical Journal}

\shorttitle{Heliospheric Compression by Supernovae}
\shortauthors{Miller \& Fields}

\graphicspath{{./}{figures/}}

\begin{document}

\title{Heliospheric Compression due to Recent Nearby Supernova Explosions}

\author[0000-0001-5071-0412]{Jesse A. Miller}
\affiliation{Department of Astronomy, University of Illinois, Urbana, IL 61801, USA}
\affiliation{Center for Advanced Studies of the Universe, University of Illinois, Urbana, IL 61801, USA}

\author[0000-0002-4188-7141]{Brian D. Fields}
\affiliation{Department of Astronomy, University of Illinois, Urbana, IL 61801, USA}`
\affiliation{Center for Advanced Studies of the Universe, University of Illinois, Urbana, IL 61801, USA}
\affiliation{Department of Physics, University of Illinois, Urbana, IL 61801, USA}

\begin{abstract}
The widespread detection of \fe60 in geological and lunar archives
provides compelling evidence for recent nearby supernova explosions
within $\sim 100$ pc around 3 Myr and 7 Myr ago.
The blasts from these explosions had a profound effect on the heliosphere.
We perform new calculations to study the
compression of the heliosphere due to a supernova blast.
Assuming a steady but non-isotropic solar wind,
we explore a range of properties appropriate
for supernova distances inspired by recent \fe60 data,
and for a 20 pc supernova proposed to account for mass extinctions at the end-Devonian period.
We examine the locations of the termination shock
decelerating the solar wind
and the heliopause
that marks the boundary between the solar wind and supernova material.
Pressure balance scaling
holds, consistent with studies of other astrospheres.
Solar wind anisotropy does not
have an appreciable effect on shock geometry.
We find that supernova explosions at 50 pc (95 pc)
lead to heliopause locations at 16 au (23 au) when the forward shock arrives.
Thus, the outer
solar system was
directly exposed to the blast, but the inner planets---including the Earth---were not.
This finding reaffirms that the delivery of supernova material to the Earth is not from the blast plasma
itself, but likely is from supernova dust grains.
After the arrival of the forward shock, the weakening supernova blast
will lead to a gradual rebound of the heliosphere,
taking $\sim100$s of kyr to expand beyond 100 au.
Prospects for future work are discussed.

\end{abstract}


\section{Introduction} \label{sec:intro}

The local neighborhood of the Sun is an ever-changing environment,
as a result of our residence in a star-forming galaxy
with a dynamic interstellar medium.
The heliosphere must therefore evolve with 
time in response \citep{muller_heliospheric_2006, muller_heliosphere_2009, frisch_suns_2006, frisch_interstellar_2011}.
Indeed, events on Galactic scales may have an impact on Earth. Terrestrial ice ages have been linked to our passage through the Galaxy's spiral arms \citep{gies_ice_2005}, and our vertical motion through the Galactic disk may have affected the cosmic ray flux on Earth and have corresponding biological signatures \citep{medvedev_extragalactic_2007}.
The Sun's passage through a dense cloud could compress the heliosphere to 1 au or smaller \citep{yeghikyan_consequences_2003, yeghikyan_effects_2004};
it has even been suggested that a recent such event could have occurred, and the
corresponding compression could have ultimately had an effect on human evolution \citep{opher_climate_2022}.
In this work, we explore the hydrodynamic effects of near-Earth supernovae, in which the blast wave moves at a  velocity much greater than the Sun's typical speed through the local interstellar medium (ISM).

There is now abundant evidence of multiple
recent near-Earth supernovae within $\lesssim$ 100 pc.
The most compelling is the discovery of live (undecayed) samples of \fe60 ($t_{1/2} = 2.6$ Myr).
This radioactive isotope is found in geological records such as ocean sediments and crusts \citep{knie_ams_1999, knie_f_2004, fitoussi_search_2008, wallner_recent_2016, ludwig_time-resolved_2016, wallner_60fe_2021}, Antarctic snow \citep{koll_interstellar_2019}, and even lunar samples returned by {\em Apollo} astronauts \citep{fimiani_interstellar_2016}.
\fe60 has also been found in cosmic rays \citep{binns_observation_2016}, which
show a low-energy excess Fe flux that could be evidence for a recent nearby source \citep{boschini_discovery_2021}. \citet{wallner_60fe_2021} also detected \pu244 coincident with the \fe60 signals. In addition, \citet{korschinek_supernova-produced_2020} reports a detection of \mn53, and \al26 has been studied but not yet  separated from the overwhelming terrigenic component \citep{feige_limits_2018}.

Further evidence of a nearby supernova comes from the proton, antiproton, and positron cosmic-ray spectra and anisotropy \citep{kachelries_cosmic_2018, savchenko_imprint_2015}, the existence of the Local Bubble \citep{smith_multiple_2001, frisch_effect_2017}, and the observed distribution of  Galactic \al26 \citep{fujimoto_distribution_2020}.
By tracing back the motion of runaway stars and neutron stars, \citet{neuhauser_nearby_2020} suggested stars in a binary progenitor system for these supernovae; similarly, \citet{tetzlaff_neutron_2013} also suggested a progenitor for the nearby Antlia supernova remnant \citep{mccullough_discovery_2002}.
The short \fe60 lifetime demands that it was produced recently
and thus nearby.  These data are consistent with a nearby supernova $\sim$3 Myr ago,
and the \fe60 abundance implies a supernova distance of $60-130\ {\rm pc}$ \citep{fry_astrophysical_2015}.

The observed \fe60 came to us in the form of dust; \fe60 ions would have been a component of the plasma that gets deflected by the heliosphere.
The dynamics of dust grains in the outer heliosphere 
have received considerable theoretical study 
\citep[e.g.,][]{belyaev_dynamics_2010, sterken_flow_2012}, especially for the present-day heliosphere.
\citet{wallis_penetration_1987} found that during the passage through a dense cloud, dust can penetrate the heliosphere to Earth with little deflection due to the heliosphere's small size.
\citet{athanassiadou_penetration_2011, fry_radioactive_2016} found that dust grains from near-Earth supernovae are typically deflected less than 1$^\circ$ by the heliosphere,
due to their high speeds $v_{\rm dust} \gg v_{\rm esc}(1 \, \rm au)=42 \ \rm km/s$ far exceeding the escape speed at 1 au.
The simulations we perform here may also contribute to our understanding of dust grain dynamics from near-Earth supernovae.

\citet{wallner_60fe_2021} has recently 
found \fe60 from a second pulse due to an earlier supernova $\sim$7 Myr ago, showing
that nearby supernovae are relatively commonplace on geological and astrophysical
timescales.  These two known events are at roughly similar distances, too far to 
cause mass extinctions of species on Earth, although possible damage to the biosphere
is an open question under study
\citep{melott_cosmic_2019, melott_supernova_2017, thomas_terrestrial_2016}.
Closer events should occur, but less frequently. With this in mind, 
\citep{fields_supernova_2020} proposed that one or more supernovae 
at $\sim 20 \ \rm pc$ could have triggered extinctions
at the end of the Devonian period 360 Myr ago, leading to observed
global ozone depletion reflecting ionizing radiation damage from the explosion.

Motivated by these data,
we consider the case of a near-Earth explosion, 
one of the most dramatic events the heliosphere can experience.
This scenario is
relevant for both the distant past and recent well-documented events.
In the aftermath of a supernova, the supernova remnant (SNR) rapidly expands outwards, sweeping
up the interstellar medium and eventually engulfing many surrounding stars. As the blast wave encounters these stars, it drives back their stellar winds, compressing their astrospheres.
We aim to study the extent of this compression as applied to own heliosphere with a suite of numerical simulations to determine the innermost distance the supernova blast penetrates in our solar system. 

To date, the only simulations of supernovae interacting with the heliosphere has been done in \citet{fields_supernova_2008},
which lays the foundations for our work here.
This earlier study examined the impact of supernovae out to at most 30 pc, closer than current estimates suggest for the 3 Myr event \citet{fry_astrophysical_2015}.  
Our work will for the first time study blasts from supernovae out to 126 pc, in line with 
the results from analyses of the \fe60 data.
We also perform detailed comparisons of our solar wind model against {\it in situ} measurements
from {\em Voyager 2} and {\em Ulysses}.  In addition, we study the scaling of the heliosphere dimensions
with the supernova blast properties in a more detailed manner.  We then use these scalings to estimate the evolution of heliosphere
compression with the arrival of the supernova shock and subsequent rebound towards its present boundary.

The structure of this paper is as follows: section \ref{sec:setup} describes the formalism and initialization of the simulations as well as expectations; section \ref{sec:simulations} presents the results of the simulations; section \ref{sec:discussion} discusses these results; and section \ref{sec:conclusion} gives concluding remarks.

\section{Simulation model} \label{sec:setup}

Our goal is to examine how our heliosphere is compressed by a supernova blast wave. Since the most recent time this occurred was $\sim$3 Myr ago, present-day observations of this phenomenon  are impossible. Therefore, we must turn to numerical simulations. We begin with the basic fluid equations, and then show that they produce a solar wind that roughly agrees with observations. Next we apply the Sedov model for a supernova remnant as the input for the blast wave. Finally we explore scaling laws for how these flows should interact.

\subsection{Fluid equations} \label{subsec:eqns}

The modern heliosphere enjoys a variety of complex physics that deviates from standard hydrodynamics, such as magnetic fields, multi-fluid flows, charge exchange, and cosmic ray propagation \citep{pauls_interaction_1995, zank_interaction_1999}.
Our goal, however, is to investigate the broad changes induced by the supernova blast.
Consequently, we do not attempt to compete with the sophisticated models that include these effects,
such as those presented in, e.g., \citet{pogorelov_three-dimensional_2004}, \citet{izmodenov_modeling_2008}, and \citet{opher_magnetized_2015, opher_small_2020}.

Indeed, some of the relevant physics for the modern heliosphere may not be applicable for the case of an incoming supernova blast. For example,
charge exchange occurs when neutral ISM atoms penetrate into the heliosphere \citep{baranov_model_1993, pauls_interaction_1997}.
As a result, the solar wind's ram pressure is weakened and the boundary between the solar wind and the present-day ISM is closer than it would be for a fully ionized ISM.
But as we will show, we do not expect a SNR to contain a large population of neutrals, in which case the effects of charge exchange are not at play.  Out to 30 au, the one-component model of the solar wind matches the observed density very well and the velocity to a difference of less than 10\% \citep[see Fig. 4.2 in][]{zank_interaction_1999},
a result we will confirm below.

To start, we assume that both solar wind and SNR flows can be adequately described with the basic equations of hydrodynamics,
\begin{eqnarray}
    \frac{\partial \rho}{\partial t} + \bm{\nabla} \cdot (\rho \bm{v}) &= 0\\
    \frac{\partial \rho \bm{v}}{\partial t} + \bm{\nabla} \cdot (\rho \bm{v v} + p) &= 0\\
    \frac{\partial E}{\partial t} + \bm{\nabla} \cdot [(E+p)\bm{v}] &= 0
\end{eqnarray}
where $p = nkT$ for an ideal gas. We use an adiabatic equation of state with $\gamma = 5/3$.

We solve these equations with the \texttt{Athena++} code \citep{stone_athena_2020}. \texttt{Athena++} is a grid-based magnetohydrodynamics framework, though we only make use of its hydrodynamics and neglect magnetic fields.
We use the built-in HLLE Riemann solver, well-suited for our case where the kinetic energy of the flow dominates.
While this solver can be diffusive, especially around contact discontinuities,
it suppresses the Carbuncle instability and the ``odd-even decoupling'' that can appear when using other solvers \citep{sutherland_numerical_2003, quirk_contribution_1994}. While performing our simulations, we do not see evidence for substantial diffusion around any contact discontinuities.

\subsection{Solar wind initialization}

The solar wind is the complex flow of gas launched from the solar corona and streaming outwards through the solar system, first predicted by \citet{parker_dynamics_1958}.
While the real solar wind varies with time according to solar activity
\citep{provornikova_plasma_2014,izmodenov_modeling_2008},
for these simulations we adopt a constant, steady outflow.
Spacecraft near Earth's orbit such as \textit{ACE} and \textit{DSCOVR} have taken an abundance of solar wind data.
Our main interest is in how the solar wind interacts far away from the Sun, so we input the wind at 1 au using a rough average of solar wind density, speed, and thermal pressure. All grid cells within 1 au are overwritten with constant values every timestep.

The real solar wind not only varies in time, but also location: it is launched differently in the plane of the solar system than towards the poles. We use data from {\em Ulysses}'s close approach in 1995  to approximate these parameters as a step function in angle, shown in Table \ref{tab:sw-input}. For comparison, we also show the values adopted in \citet{fields_supernova_2008}, which is spherically symmetric.

\begin{table*}[htb]
    \centering
    \caption{Solar wind input values at 1 AU}
    \label{tab:sw-input}
 \hspace{-2cm}
 \begin{tabular}{ccccc}
    \hline \hline
    Region & Density & Velocity & Ram pressure ($\rho v^2$) & Thermal pressure\\
    & (g cm$^{-3}$) & (km s$^{-1}$) & (erg cm$^{-3}$) & (erg cm$^{-3}$)\\
    \hline
    \multicolumn{1}{r}{\citet{fields_supernova_2008} comparison, global} & 1.02 $\times$ 10$^{-23}$ & 464 & 2.19 $\times$ 10$^{-8}$ & 2.00 $\times$ 10$^{-10}$\\
    \multicolumn{1}{r}{equatorial, $|\theta| < 25\degree $} & 1.06 $\times$ 10$^{-23}$ & 434 & 2.00 $\times$ 10$^{-8}$ & 6.68 $\times$ 10$^{-10}$\\
    \multicolumn{1}{r}{polar, $|\theta| > 25\degree $} & 4.18 $\times$ 10$^{-24}$ & 643 & 1.73 $\times$ 10$^{-8}$ & 9.91 $\times$ 10$^{-10}$\\
    \hline \hline
 \end{tabular}
 \end{table*}

While a reasonable wind initialization is made across the grid at the start of the simulation, the solar wind is allowed to relax to its steady-state profile before the supernova blast is introduced.

\begin{figure}[!htb]
    \centering
    \includegraphics[width=0.47\textwidth]{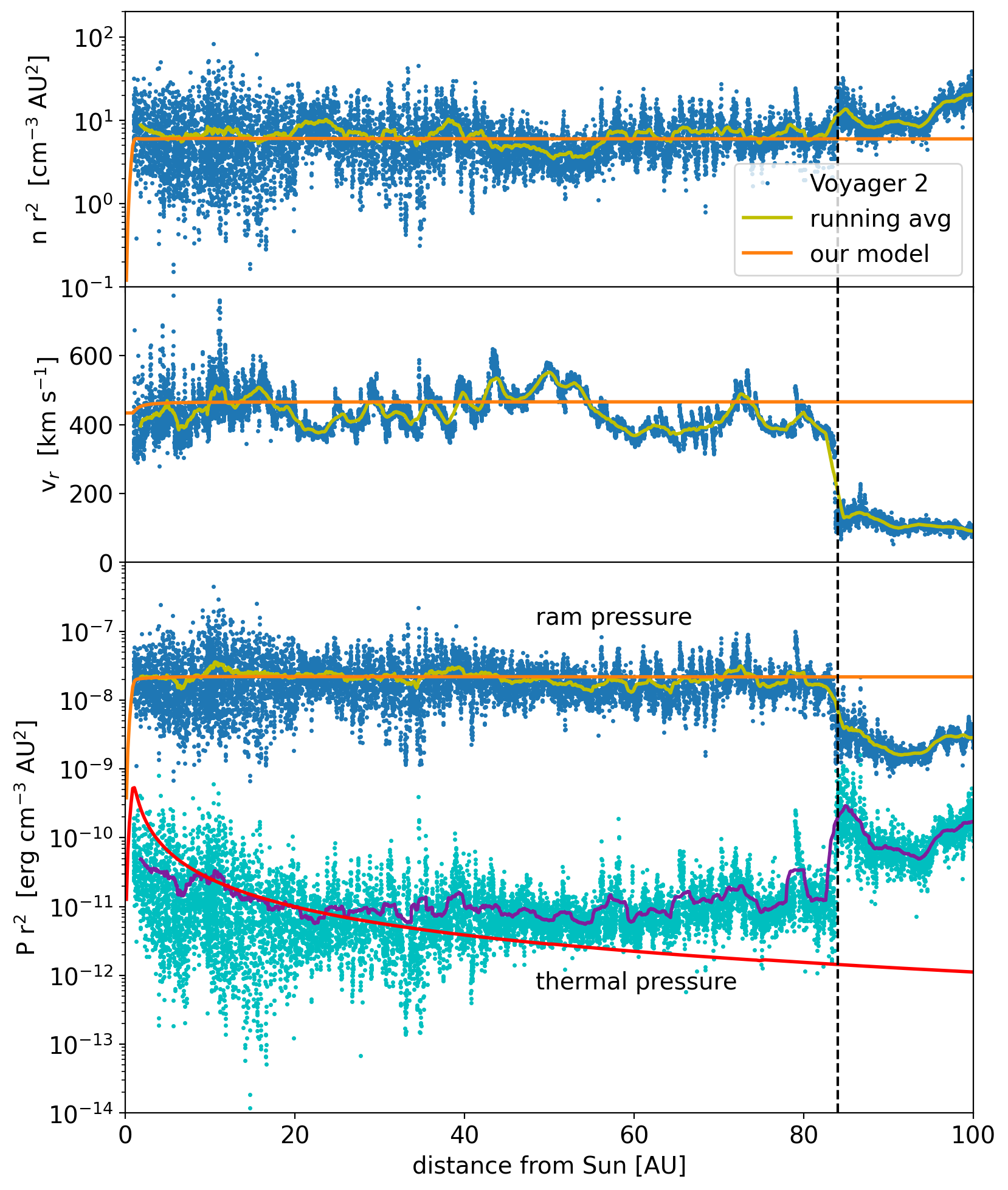}
    \caption{Comparison of our equatorial solar wind quantities (orange/red line) to daily averages of measurements made by {\em Voyager 2} (blue/cyan points) and a 240-day running average of {\em Voyager 2} (yellow/violet line). The black dashed line shows where {\em Voyager 2} crossed the termination shock, beyond which our model is not expected to match the data. The top panel shows density (scaled by $r^2$), the middle panel shows velocity, and the bottom panel shows ram and thermal pressure (scaled by $r^2$).}
    \label{fig:sw_init_V2}
\end{figure}

With our steady one-fluid model, we must check to ensure our simulated solar wind still accurately represents the observed heliosphere. To this end, we compare our relaxed solar wind in the equatorial region to {\em Voyager 2} plasma data. Figure \ref{fig:sw_init_V2} shows density, velocity, and thermal and ram pressures of both our model and {\em Voyager 2} data. A 240-day running average of {\em Voyager 2} data is also shown. Our model tracks {\em Voyager 2}'s density, velocity, and ram pressure well, though not accounting for temporal variations. The thermal pressure clearly has a different profile. We attribute this in part to the assumption of the adiabatic evolution of our model, whereas the real solar wind is not adiabatic. Since the thermal pressure approximately matches in the region from 5--40 au, we consider it to be sufficiently accurate in the most relevant region. Furthermore, the thermal pressure is still several orders of magnitude below the ram pressure, which will dominate the large-scale structure of the simulations.

\begin{figure}[!htb]
    \centering
    \includegraphics[width=0.47\textwidth]{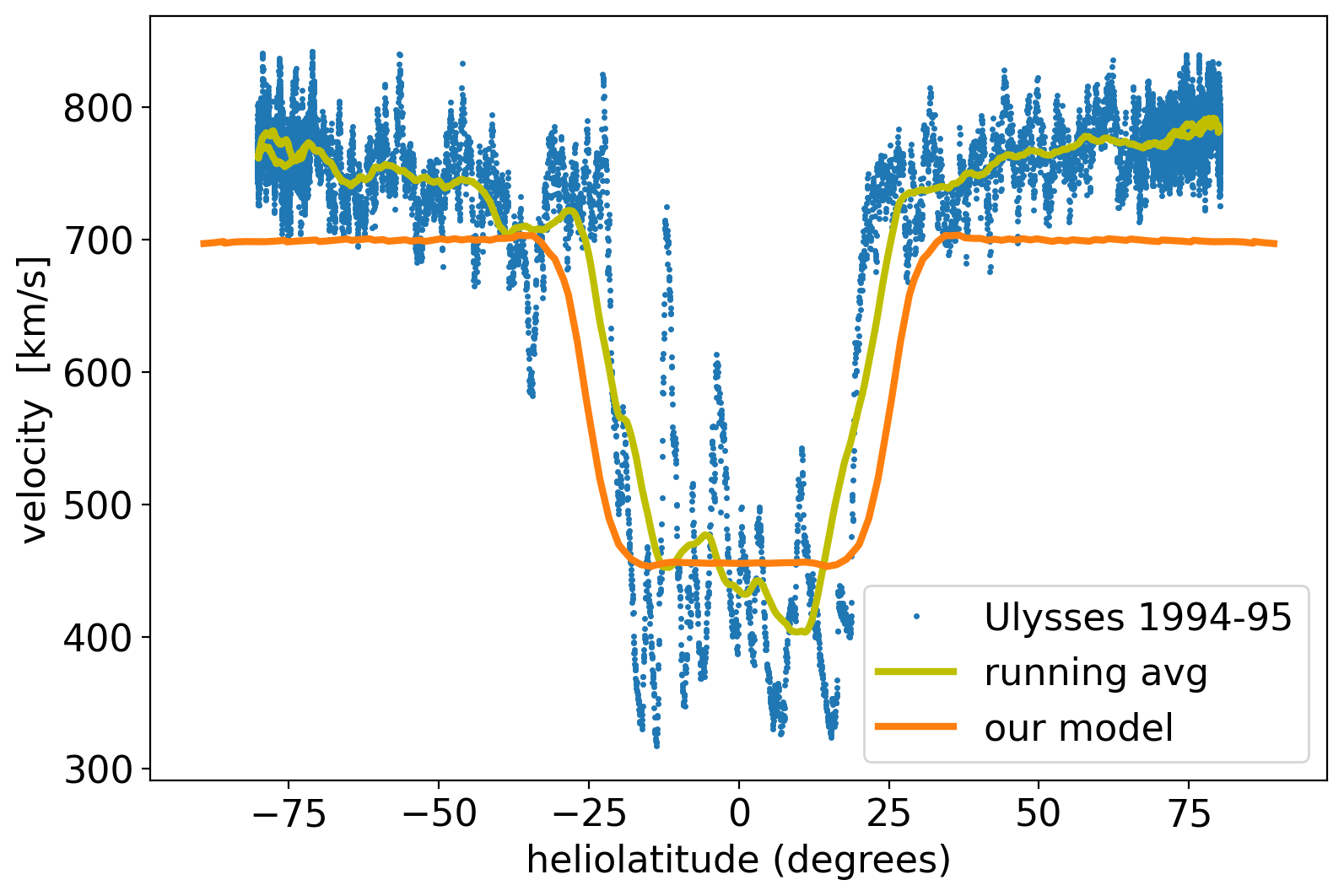}
    \caption{Comparison of our solar wind velocity to measurements made by {\em Ulysses} during its close approach of the solar minimum of 1995. We use the same color scheme as in Figure \ref{fig:sw_init_V2}.
    Solar wind velocity does not change appreciably with distance (Fig.~\ref{fig:sw_init_V2}) but
    simulations are for a distance of 2 au; data follow the {\em Ulysses} elliptical orbit and sample
    distances from 1.3 to 2.8 au.}
    \label{fig:sw_init_Uly}
\end{figure}

In Figure \ref{fig:sw_init_Uly} we plot the velocity as a function of heliolatitude during {\em Ulysses}' close approach. Although initially a polar step function, the abrupt change in velocity is slightly smoothed out as the wind propagates. Data from our model is taken from a distance of 2 au in order to allow the solar wind time to relax and give a better description of the distant solar wind than the input step function. The largest discrepancy here is due to temporal variability, which we do not model. The averages over time are a close match.

Currently, it is an open question as to where the supernova 3 Myr ago exploded, and therefore which direction the blast wave would come from. Proposed clusters include the Sco-Cen association \citep{benitez_evidence_2002} and the Tuc-Hor association \citep{mamajek_pre-gaia_2015, hyde_supernova_2018}. \citet{sorensen_near-earth_2017} traced back nearby stellar clusters and statistically examined which ones were most likely to produce a nearby supernova explosion.
In addition, this supernova could have been a companion to a previous star that exploded, and could have been flung outwards with orbital velocity and exploded away from its home cluster. Given these uncertainties in location, we do not assume a single place of the supernova, but instead examine the effect of solar wind orientation with respect to the supernova.

\begin{figure*}[!htb]
\centering
\gridline{\fig{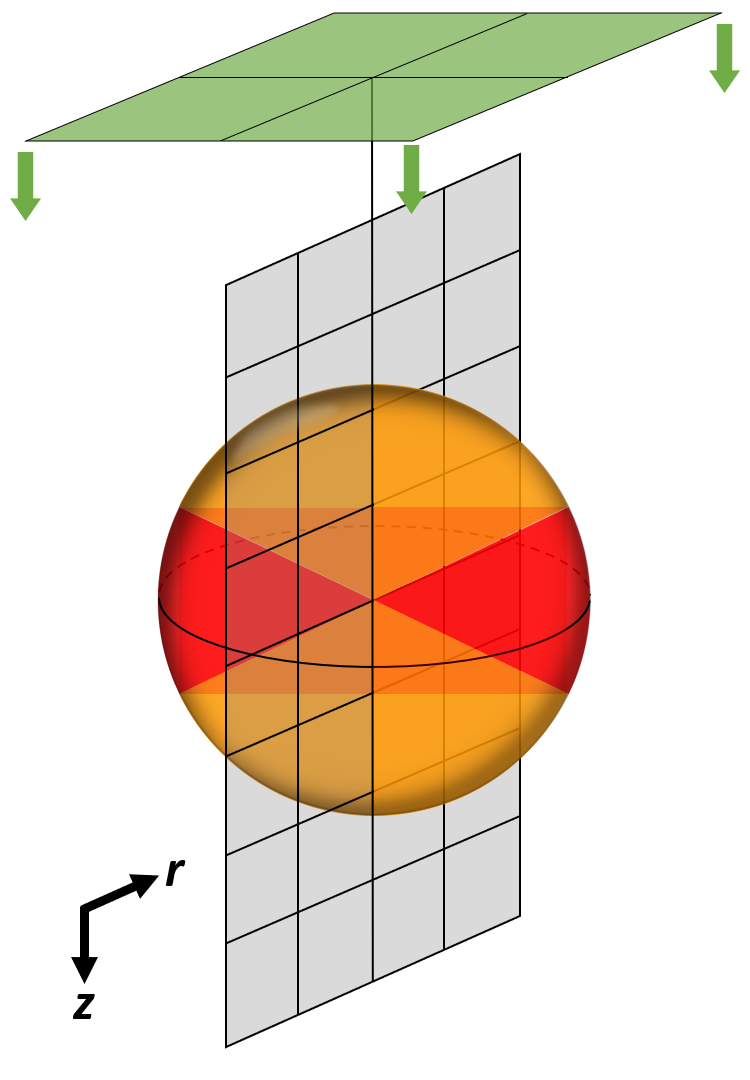}{0.3\textwidth}{(a)}
          \fig{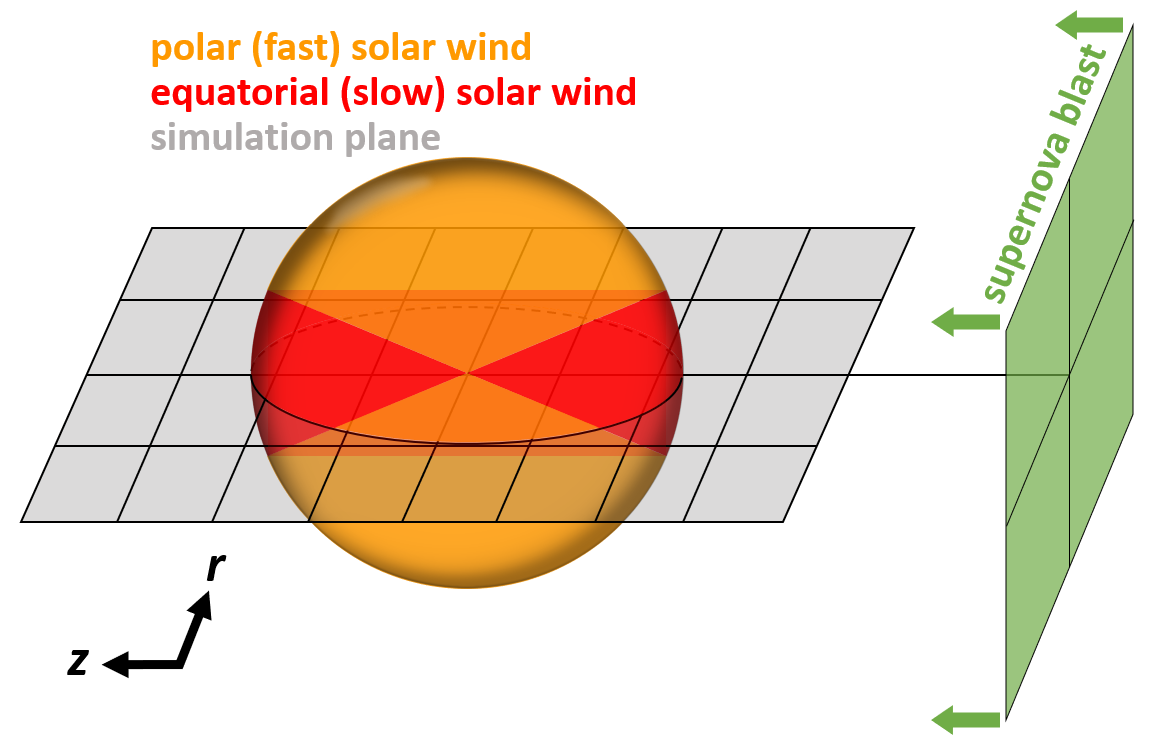}{0.5\textwidth}{(b)}
          }
    \caption{Schematic of the two orientations. The equatorial solar wind (red) is slow and dense, while the polar solar wind (yellow) is fast and sparse. The gray region shows the 2D plane of the simulation domain.
    (a) In the polar orientation, the simulated wind is polar along the $\mathbf{z}$ direction and equatorial along $\mathbf{r}$.
    (b) In the equatorial orientation, the entire solar wind in the simulation is equatorial, resulting in an isotropic wind.}
    \label{fig:orientation}
\end{figure*}

In order to increase computational efficiency, we perform our simulations in 2D cylindrical coordinates in the $r$-$z$ plane, with rotational symmetry imposed about the $z$ axis. 
We use two different orientations depending on whether the blast wave enters orthogonal or parallel to the plane of the solar system.
Figure \ref{fig:orientation} shows a schematic geometric interpretation of these orientations.
When the blast comes perpendicular to the axis of rotation as shown in panel (a), 
the rotational symmetry of our simulation captures the solar wind behavior, which
we model with the step function in angle described above; this is the polar orientation.
When the blast comes in along the plane (striking the outer planets' orbits first) as in panel (b), we use a spherically symmetric wind; we refer to this as the equatorial orientation.
Both orientations have similar ram pressures, so we do not expect great differences in heliospheric structure between the orientations.

\begin{figure}[!htb]
    \centering
    \includegraphics[width=0.47\textwidth]{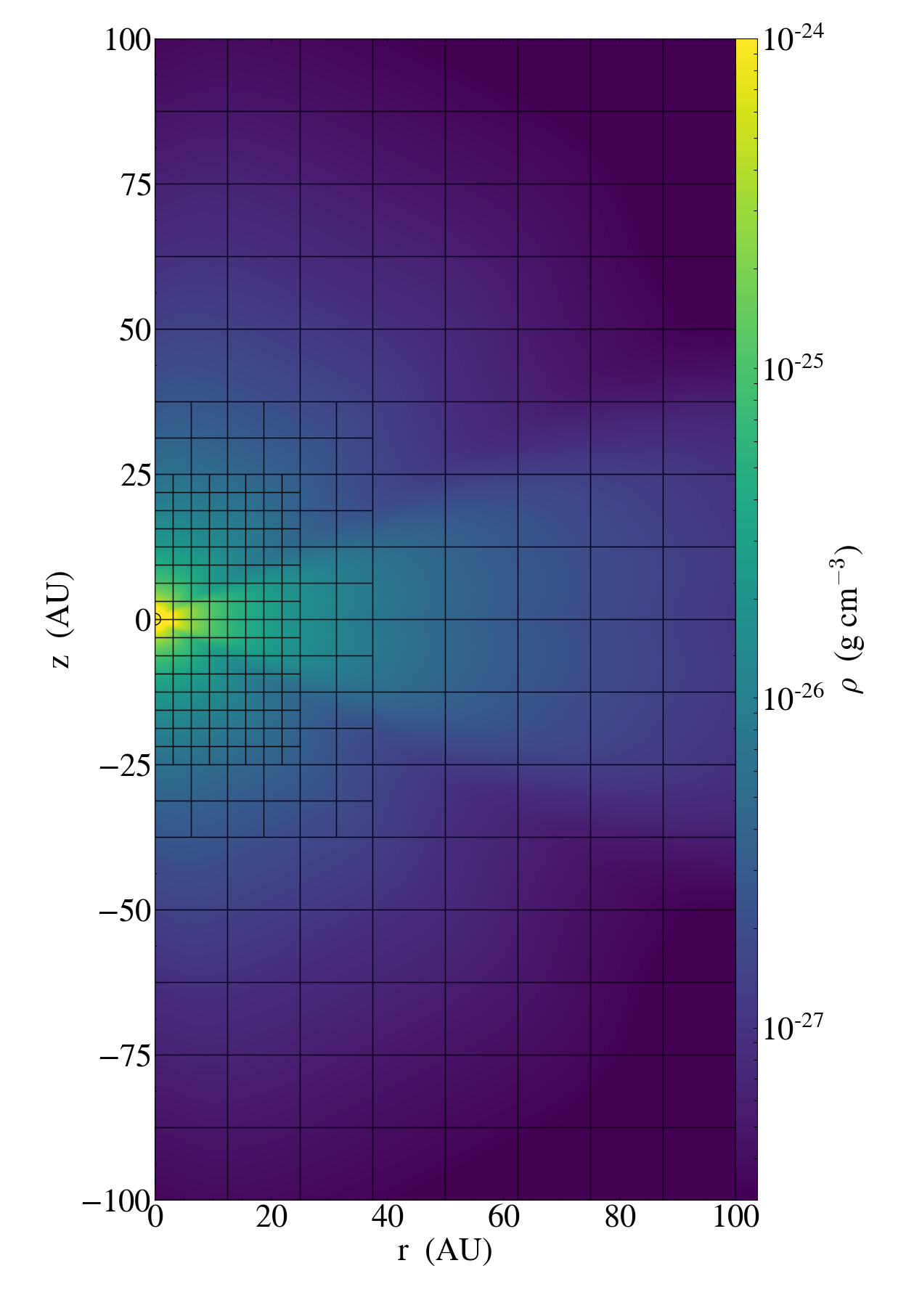}
    \caption{Density plot of the initialized, relaxed solar wind in the polar orientation. Color shows density on a log scale. The black grid shows \texttt{Athena++} meshblocks.}
    \label{fig:sw_init_A++}
\end{figure}

Our base mesh resolution is 256$\times$512 cells for all but the comparison to \citet{fields_supernova_2008}, which is 1024$\times$1536. If kept uniform across the grid, several of our larger simulations would make the region within 1 au only a few cells across, resulting in a poorly-defined flow. To ensure a smooth spherical outflow, we refine the solar wind injection region for the entirety of the simulation (known as static mesh refinement, SMR). The amount of refinement depends on the outer boundaries of the mesh. Figure \ref{fig:sw_init_A++} shows the the relaxed solar wind and corresponding meshblocks in the polar orientation. Each refinement level increases resolution by a factor of 2.

\subsection{Supernova blast initialization}

Supernova remnants evolve over time. After the initial free expansion phase, supernova remnant morphology roughly follows a Sedov-Taylor profile \citep{sedov_propagation_1946, taylor_formation_1950} for much of its evolution.
The remnant expands quasi-spherically over many parsecs until thermal emission becomes important.
For SNRs several tens of parsecs across, a spherical forward shock is well-approximated as plane wave on the scale of the solar system. Since the most important parameter in our simulations is the distance to the supernova, $R_{\rm SN}$, we invert the usual Sedov equation for distance to solve for time as
\begin{equation}
    \label{eq:tarrive}
    t = \sqrt{\frac{R_{\rm SN}^5 \rho_0}{\beta^5 E_{\rm SN}}} \\
    = 27 \ {\rm kyr} \ \pfrac{n_0}{0.01 \ \rm cm^{-3}}^{1/2} \pfrac{R_{\rm SN}}{50 \ \rm pc}^{5/2},
\end{equation}
where $R_{\rm SN}$ is the radius of the remnant, $\rho_0$ is the ambient medium density, $\beta$ is a numerical factor of 1.1517 for the typical $\gamma = 5/3$, and $E_{\rm SN}$ is the energy of the supernova, taken to be $10^{51}$ ergs.
By applying the Rankine-Hugoniot conditions, we find the density, velocity, and thermal pressure of the gas immediately post-shock, which are
\begin{align}
    \rho_1 &= \rho_0 \frac{\gamma +1}{\gamma -1}  \label{eqn:sedov_rho}\\
    v_1 &= \frac{2}{\gamma+1} v_s  \label{eqn:sedov_v}\\
    p_1 &= \frac{2}{\gamma+1} \rho_0 v_s^2 ,  \label{eqn:sedov_p}
\end{align}
where the subscripts 0 and 1 indicate the ambient medium and immediate post-shock gas, respectively. $\gamma$ is the adiabatic index, and $v_s$ is the shock velocity, given by
\begin{eqnarray}
    v_s & = &  \frac{2}{5} \frac{R_{\rm SN}}{t} = \frac{2}{5}\sqrt{\frac{E_{\rm SN}}{\beta^5 \rho_0}} R_{\rm SN}^{-3/2} \\
    & = & 730 \ {\rm km/s} \ \pfrac{n}{0.01 \ \rm cm^3}^{-1/2} \pfrac{R_{\rm SN}}{50 \ \rm pc}^{-3/2}.
\end{eqnarray}

Our current local interstellar environment is dominated by the Local Bubble, a region of low-density carved out by multiple supernovae \citep{smith_multiple_2001}.
While the density varies greatly with location, we use an average value of $n=0.005\ {\rm cm}^{-3}$, rather than a typical Galactic ISM.
The earliest supernovae to explode did so in a denser environment, and later ones encounter the swept-out low-density region from the past explosions.
\citep{fuchs_search_2006} estimated that 14-20 supernovae have exploded in the Local Bubble, and \citep{schulreich_numerical_2017} and \citet{breitschwerdt_locations_2016} performed hydrodynamic simulations showing how 16 supernovae can generate this environment.
We will show that the ambient density has no effect on the distance of closest approach in our solar system, a feature of using the Sedov model.
But the low density  of the medium extends the duration of the Sedov phase, 
which should be a reasonable approximation for the supernova distances of interest here.

Furthermore, we can calculate the temperature of the post-shock material assuming an ideal gas. This yields a remarkably high temperature of
\begin{eqnarray}
    T &=& \frac{2 \beta^5}{25k_B (\gamma+1)} \frac{E_{\rm SN}}{r_s^3 n_1}\\
    T &=& 1.5 \times 10^6\ \left( \frac{R_{\rm SN}}{\rm 100\ pc} \right)^{-3} \left( \frac{n_0}{0.01\ \rm cm^{-3}} \right)^{-1}\ {\rm K}.
\end{eqnarray}
For the low density observed in the Local Bubble, this model predicts that the temperature post-shock will remain over $10^6$ K for 100 pc, the approximate size of the Local Bubble.
In the past, it was accepted that Local Bubble consisted of a large hot component with \mbox{$T \sim 10^6$ K}, though such assumptions have recently been challenged \citep{welsh_trouble_2009, linsky_could_2021}.
Despite the hot environment of supernova remnants, the quick formation of dust and even molecules in SN 1987A \citep{matsuura_alma_2017} suggest that some component of the remnant may be neutral.
We do not account for multi-fluid hydrodynamics here, instead interpreting the Sedov equations to indicate complete ionization at the forward shock.

The most important parameter for the supernova is the distance, which directly affects the strength of the blast wave, with ram pressure scaling as $P_{\rm ram} \sim E_{\rm SN}/R_{\rm SN}^3$. The supernova 3 Myr ago is estimated to have occurred somewhere between 60-130 pc away \citep{fry_astrophysical_2015}. Recently, \citet{fields_supernova_2020} proposed that a supernova at $\sim 20$ pc could have contributed to biological extinctions at the end-Devonian period. To cover this range of distances, we use supernova distances of 25.3, 50.0, 63.3, 75.9, 94.9, 110.0, and 126.5 pc. \citet{looney_radioactive_2006} proposed that a very nearby supernova within $\sim 1\ \rm pc$ exploded in the early stages of the solar nebula; we limit our simulations to the fully-formed solar system and so do not consider this case here.

From a single point in space, SNRs weaken over tens of thousands of years. Our simulations cover at most the first few years of the initial blast; this timescale is a multiple of the $\sim 0.5 \ \rm yr$ crossing time for a flow of $100 \ \rm km/s$ to travel $10 \ \rm au$.
Therefore, we do not allow the remnant to weaken during our simulation. Since we are primarily interested in the closest approach of the blast wave in our solar system, we restrict ourselves to modelling the strongest part of the blast, the forward shock.  See \S \ref{sect:weaken}
for discussion of the behavior over longer timescales.

The supernova blast is implemented as a boundary condition in \texttt{Athena++}. We first let the solar wind evolve until it has reached a steady state across the entire domain. Then the boundary condition at $-z_{\rm max}$ is changed to be the incoming supernova blast, flowing in the $+z$
direction.
This new boundary condition is kept constant for the duration of the simulation. All times shown in our figures define $t=0$ as the introduction of the blast.

\subsection{Expected heliosphere structure and stagnation distance}

Our work draws upon insights from the many studies of
the {\em present-day} heliosphere's interaction with the very local ISM \citep[e.g.,][]{opher_small_2020, frisch_interstellar_2011}. In addition, several studies of stellar winds interacting with the ISM have been carried out, both observational and numerical \citep{meyer_3d_2021, henney_bow_2019, kobulnicky_comprehensive_2016}.
These studies show that the solar wind-ISM interaction region consists of three main features: the termination shock (TS), where the solar wind slows to a subsonic velocity, the heliopause (HP), where the solar wind and in-flowing ISM meet, and the bow shock (BS), where the ISM transitions from supersonic to subsonic.
In the modern day, the {\em Voyager 1} and {\em Voyager 2} missions have passed the HP at distances of 121.6 and 119.0 au, respectively \citep{burlaga_magnetic_2019}. There is increasing evidence that there is no BS \citep{mccomas_heliospheres_2012}, indicating that the Sun's motion through the ISM is barely subsonic. This is not the case for a very rapidly-moving supernova blast.

We expect that the closest approach of the blast wave, directly on-axis, will be the point of pressure balance. (As we will show in section \ref{sec:discussion}, pressure balance is instead an excellent predictor of the TS rather than the HP.) This point, also called the stagnation distance, is relevant for astrospheres and their bow shocks \citep{wilkin_exact_1996, comeron_numerical_1998}. While commonly written as a function of a star's mass loss rate as
\begin{equation}
    r_{\rm stag} = \sqrt{\frac{\dot{M}v_w}{4\pi \rho_0 v_*^2}},
\end{equation}
where $v_w$ is the stellar wind velocity and $v_*$ is the speed of the star relative to the ISM \citep{comeron_numerical_1998}, our model of the Sun has both an equatorial and polar region with slightly different mass loss rates. In order to relate the stagnation distance to our input parameters, we write it as a balance of thermal and ram pressure between the Sun and the supernova remnant. This stagnation distance is
\begin{equation}
    \label{eqn:r_stag}
    r_{\rm stag} = \sqrt{\frac{P_{\rm sw} + \rho_{\rm sw} v_{\rm sw}^2}{P_{\rm SNR} + \rho_{\rm SNR} v_{\rm SNR}^2}} \ {\rm au}
\end{equation}
where solar wind properties are evaluated at 1 au.
\citet{fields_supernova_2008} found good agreement with this (for the TS) for very nearby supernovae. Given that the SNR gas parameters depend completely on distance, $r_{\rm stag}$ could equivalently be written in terms of the Sedov supernova distance as
\begin{align}
    r_{\rm stag} &= \left( \frac{25(\gamma-1)}{8 \beta^5} \frac{R_{\rm SN}^3}{E_{\rm SN}} \left( P_{\rm sw} + \rho_{\rm sw} v_{\rm sw}^2 \right) \right) ^{1/2} \\
    &= A \left( \frac{R_{\rm SN}}{100\ {\rm pc}} \right)^{3/2} \left( \frac{E_{\rm SN}}{10^{51}\ {\rm erg}} \right)^{-1/2} \label{eqn:rstag_num}
\end{align}
where $A = 24.97\ (23.50)$ au in the equatorial (polar) orientation. This final equation allows us to write the stagnation distance solely in terms of the supernova distance, given an explosion energy. 

Qualitatively, different supernova distances should yield the same large-scale heliospheric shape and features among the simulations. Nearby supernovae will produce a much larger velocity than distant ones, which may affect the production of Kelvin-Helmholtz instabilities that form on the HP.

\section{Results} \label{sec:simulations}

We run 13 simulations, numbered with respect to supernova distance. Model 1 is a comparison to model 12 of \citet{fields_supernova_2008} for a 20 pc supernova. Here, the solar wind follows their parameters, rather than the updated ones for the rest of our models. Models 2-4 are for a 25.3 pc and 50 pc supernova in the equatorial and top-down orientations.
Models 5a, 5b, and 5c are all a 63.3 pc supernova in the equatorial orientation, but with higher ambient medium densities representing a more dense Local Bubble before multiple supernovae carved it out. Models 6-11 are for supernovae at larger distances in both equatorial and polar orientations.
A summary of the results is given in Table \ref{tab:sim-results}, which lists the initial conditions for the supernova and the distances of closest approach for the TS, HP, and BS. The location of the BS is not stated for the cases where it has retreated off the grid domain.

\begin{table*}[htb]
    \centering
    \caption{Supernova-heliosphere collision simulation results}
    \label{tab:sim-results}
    \hspace{-2cm}
    \begin{tabular}{ccccccccccc}
    \hline \hline
    Label & \# of SMR & Orientation$^*$ & $R_{\rm SN} $ & $\rho_{\rm SNR} $ & $v_{\rm SNR} $ & $p_{\rm SNR} $ & $r_{\rm stag} $ & $r_{\rm TS}$ & $r_{\rm HP}$ & $r_{\rm BS}$ \\
    & Levels & & (pc) & (g cm$^{-3}$) & (km s$^{-1}$) & (erg cm$^{-3}$) & (au) & (au) & (au) & (au)\\
    \hline
    1$^\dag$    &0 &E  &20.0   &6.40e-25  &688   &1.00e-9  &2.34  &2.47  &3.37  &8.90\\
    2           &1 &E  &25.3   &3.34e-26  &2141  &5.11e-10 &3.18  &3.43  &4.75  &14.47\\
    3           &1 &P  &25.3   &3.34e-26  &2141  &5.11e-10 &2.99  &3.08  &3.60  &12.25\\
    4           &1 &E  &50.0   &3.34e-26  &771   &1.99e-10 &8.83  &9.72  &15.97 &---\\
    5a          &1 &E  &63.3   &3.34e-26  &541   &3.26e-11 &12.56 &13.72 &23.88 &---\\
    5b$^\ddag$  &1 &E  &63.3   &1.34e-25  &271   &3.26e-11 &12.56 &13.82 &23.97 &---\\
    5c$^\ddag$  &1 &E  &63.3   &6.68e-25  &121   &3.26e-11 &12.56 &13.82 &22.61 &38.18\\
    6           &1 &P  &63.3   &3.34e-26  &542   &3.26e-11 &11.82 &12.94 &18.90 &39.94\\
    7           &1 &E  &75.9   &3.34e-26  &412   &1.89e-11 &16.51 &17.53 &29.98 &---\\
    8           &2 &P  &94.9   &3.34e-26  &295   &9.68e-12 &21.72 &22.31 &23.10 &64.26\\
    9           &2 &E  &110.0  &3.34e-26  &236   &6.22e-12 &28.80 &30.76 &44.63 &86.33\\
    10          &2 &P  &110.0  &3.34e-26  &236   &6.22e-12 &27.11 &26.07 &27.05 &80.08\\
    11          &2 &E  &126.5  &3.34e-26  &192   &4.09e-12 &35.53 &37.21 &52.15 &97.27\\
    \hline \hline
 \end{tabular}
 \parbox{0.8\textwidth}{${}^{*}$ E = equatorial, P = polar orientation. Solar wind parameters for these are given in Table \ref{tab:sw-input}.\\
 \parbox{0.8\textwidth}{$^\dagger$ \citet{fields_supernova_2008} model 12 comparison, different solar wind parameters.}\\
 \parbox{0.8\textwidth}{$^\ddag$ Models 5b and 5c use ISM densities of $n=0.02$ and $0.1$ cm$^{-3}$, respectively.}}
\end{table*}

\begin{figure*}[!htb]
    \centering
    \begin{interactive}{animation}{animations/SWxSN_95pc_P.mp4}
    \gridline{\fig{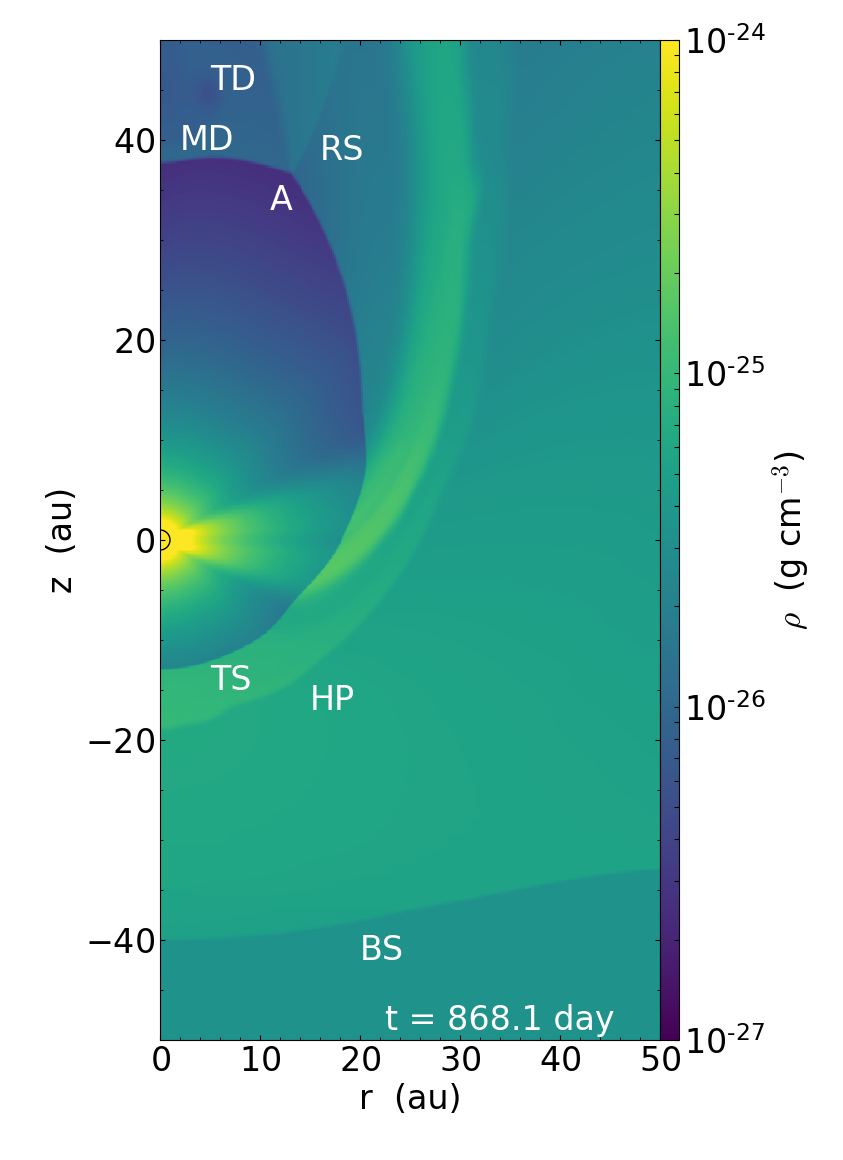}{0.32\textwidth}{(a)}
          \fig{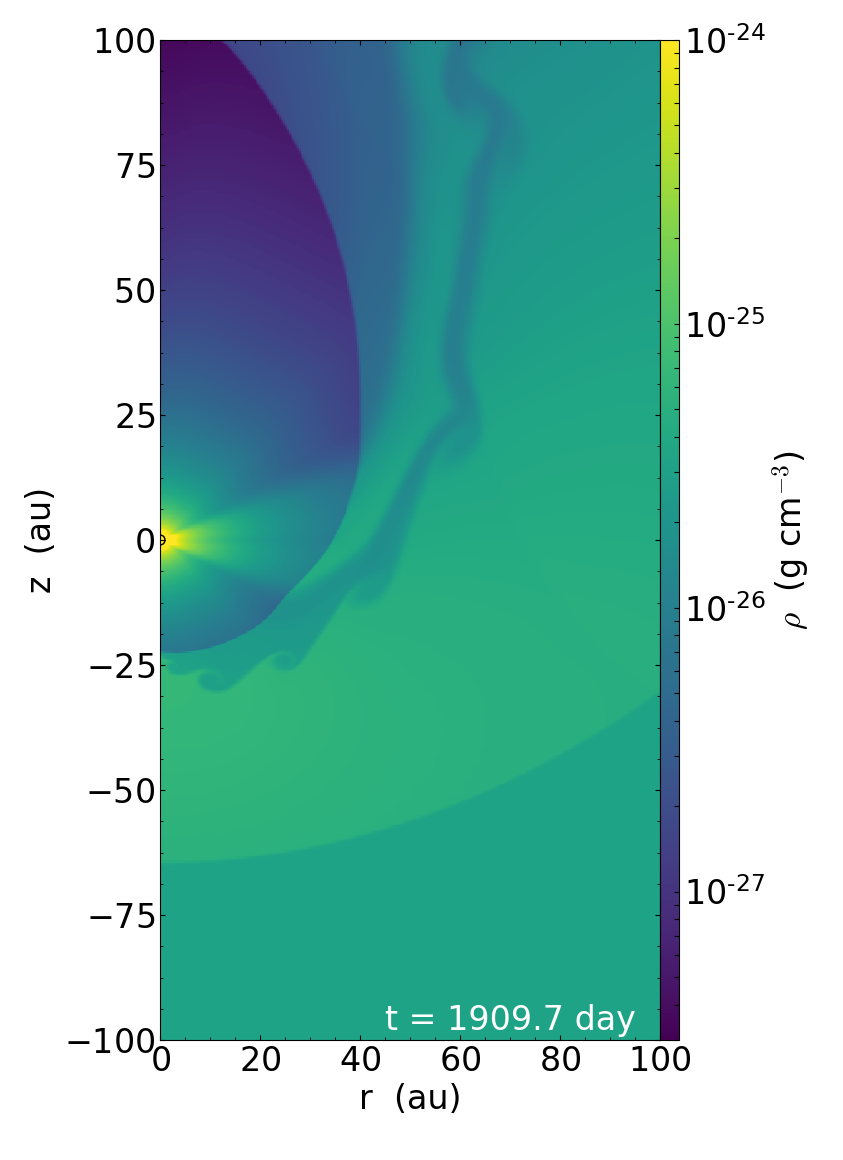}{0.32\textwidth}{(b)}
          \fig{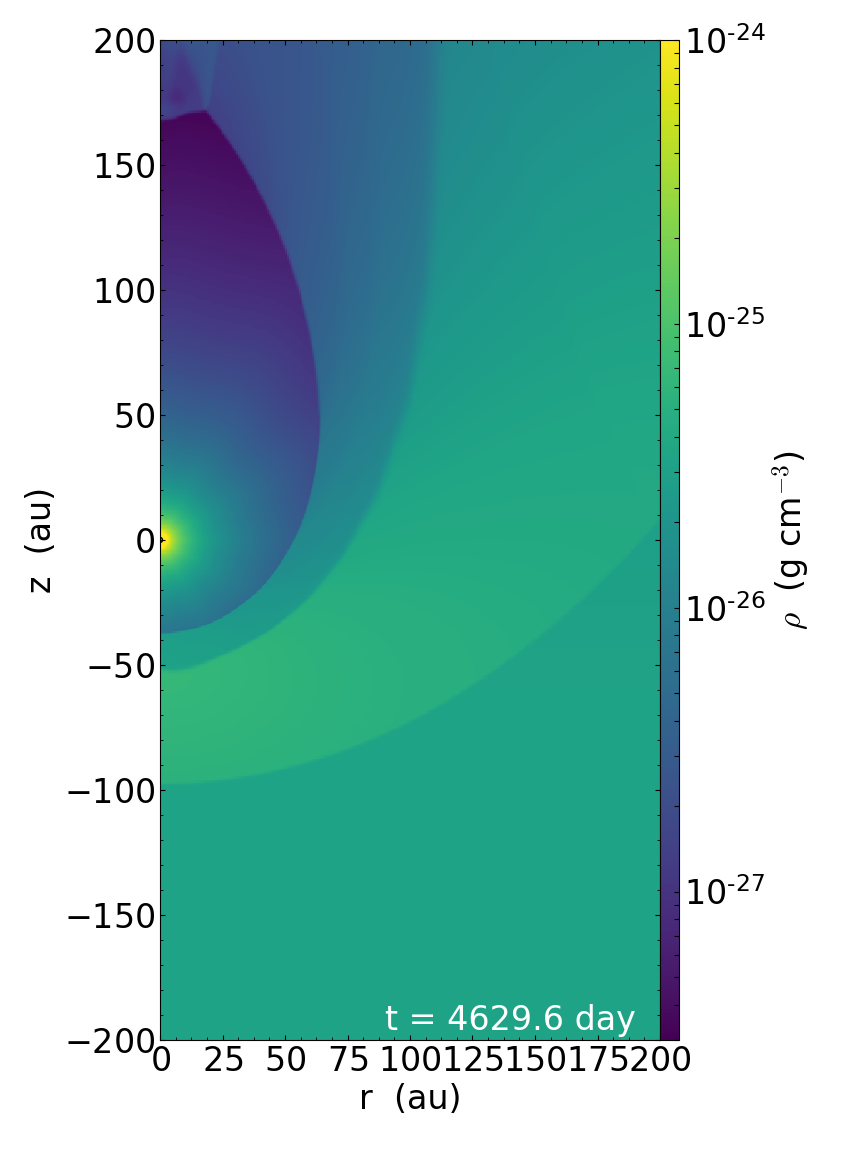}{0.32\textwidth}{(c)}
          }
    \end{interactive}
    \caption{Density plots of models 6, 8, and 11, from left to right. Note the different distance and color scales. Both (a) and (b) are in the polar orientation and (c) is in the equatorial orientation. Labelled features in (a) use the abbreviations from \citet{baranov_model_1993} and correspond to the bow shock (BS), heliopause (HP), termination shock (TS), Mach disk (MD), secondary tangential discontinuity (TD), reflected shock (RS), and the point at which the TS splits (A). An animated version of (b) is available online showing the arrival of the blast wave into the solar system at time $t=0$ and the subsequent compression of the heliosphere until the time of the displayed frame.}
    \label{fig:sims}
\end{figure*}

In Figure \ref{fig:sims} we show three density plots from the last timestep of models 6, 8, and 11. We see the qualitative expected upwind structure including the TS, HP, and BS. Any turbulence happens across the HP, seen most easily in model 8. The high-density equatorial solar wind gets bent back and is incorporated into the rest of the flow. It does not appear to be a source of turbulence. Instead, the HP drives Kelvin-Helmholtz instabilities.

Figure~\ref{fig:sims} and all of our simulations show that for supernova distances consistent with recent \fe60 measurements, the closest approach of the blast is $> 10 \ \rm au$ away. 
Certainly the blast can only arrive at 1 au for supernovae at extinction-level distances. But \citet{fry_astrophysical_2015} 
showed that 
\fe60 abundances measured in terrestrial and lunar archive over the past 10 Myr imply a supernova distance of 50-100 pc.  
This reaffirms the conclusions of
\citet{fields_supernova_2008} that
these radioisotopes must arrive in the form of dust that can decouple from the blast at the supernova-solar wind interface.  The dust must then travel $\sim 10$ au, which requires large or fast grains to avoid repulsion from the solar light pressure \citep{athanassiadou_penetration_2011,fry_radioactive_2016}.

\subsection{Comparison to previous work}

In model 1, we use the same input parameters as \citet{fields_supernova_2008} model 12 in order to compare our two codes. We find excellent agreement with the overall structure of the heliosphere and the locations of the TS and HP upwind. The largest difference between the two is the amount of Kelvin-Helmholtz instabilities at the HP present in the previous work. We attribute this abundance to the use of adaptive mesh refinement in the previous work. This refinement was not used here in favor of static mesh refinement that resolved the solar wind at 1 au over much larger scales. While further refinement is needed to study instabilities and downwind mixing in detail, the two results are otherwise consistent.

\subsection{Time-dependent features}

A characteristic timescale for these simulations is the time for the blast wave to cross the simulation domain, $t_{\rm cross}$. When $t \sim t_{\rm cross}$, the heliosphere directly upwind is in its most compressed state. Over multiple $t_{\rm cross}$, several processes continue to shape the heliosphere: Kelvin-Helmholtz instabilities along the HP, rebounding of the HP and BS, and downwind TS splitting.

The HP is a tangential discontinuity formed by solar wind and supernova fluids flowing along a surface of contact with a velocity tangential to this surface. This condition is ripe for Kelvin-Helmholtz instabilities, as we see in our simulations. Spiral features begin close to the axis of symmetry and grow as they are pushed downwind. The appearance of these instabilities is, in part, due to the resolution of our simulations, as high-resolution runs promote more instabilities and greater mixing. Since our goal is primarily to determine the innermost penetration of the SNR, we do not extend additional layers of resolution to the Kelvin-Helmholtz ripples.

\subsection{Discontinuity locations}
Tracking the location of the discontinuities (TS, HP, and BS) over time can be used to determine their stability. The Mach number $M$ is an excellent indicator of the location of strong shocks. Both the solar wind and blast wave are initially supersonic and must decrease below $M=1$ in order to interact head-on along the axis of symmetry. We note that as the blast first enters the domain, the BS does not immediately develop. Instead, there is a smooth gradient over $M$ until it sharpens into a single location after approximately $t \sim t_{\rm cross}$.

\begin{figure}[!htb]
    \centering
    \includegraphics[width=0.47\textwidth]{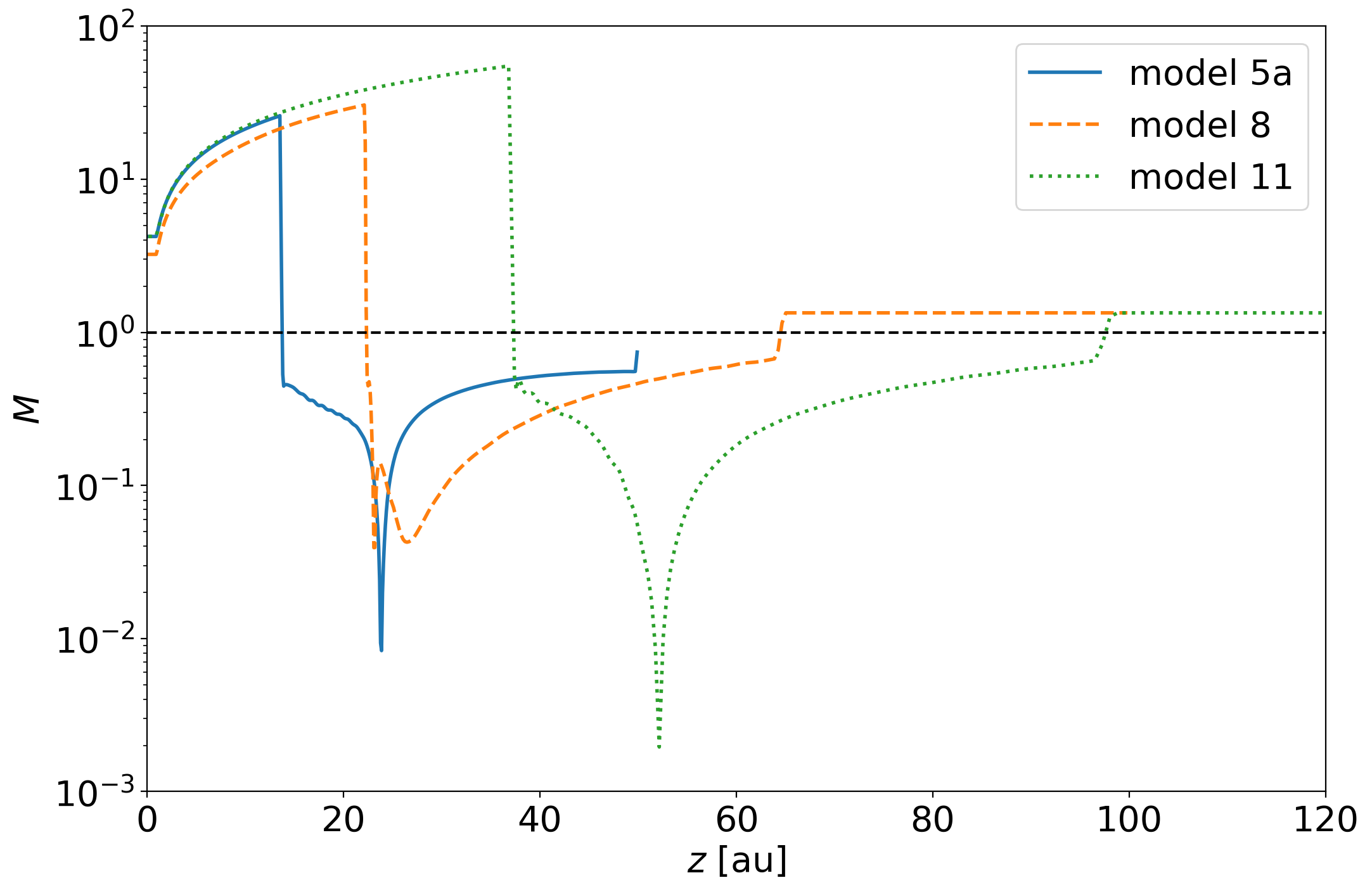}
    \caption{Profile plot of Mach number along the $z$-axis starting at the Sun and extending towards the upwind ($-z$) direction for models 5a, 8, and 11 (solid blue, dashed orange, and dotted green lines, respectively). The horizontal dashed line is at $M=1$.}
    \label{fig:mach_profile}
\end{figure}

Figure \ref{fig:mach_profile} shows the Mach number along the $z$-axis for models 5a, 8, and 11. In this plot, the locations of the TS and BS are clear, marked by the near-vertical jump in the Mach number as it crosses $M=1$. In 5a, the BS has receded out of the simulation frame, so is only marked by a small uptick at the domain boundary.

The HP is located where the two flows meet, which should be near-zero velocity along this axis. Models 5a and 11 clearly show this sharp downwards spike. The HP in model 8 is more difficult to locate. In this simulation, Kelvin-Helmholtz instabilities form close to the axis of symmetry. These ripples cause the location of the HP to change each frame, even after several $t_{\rm cross}$ have passed.
We attribute this to the polar orientation of both models: the polar solar wind is lower density, and the greater density contrast appears to promote the growth of Kelvin-Helmholtz instabilities close to the axis of symmetry. Nonetheless, we still use the minimum value for $M$ as the HP with the understanding that this value has some error of $\sim 10 \%$ for models 8 and 10.

\subsection{Discontinuities over time}

\begin{figure}[!htb]
    \centering
    \includegraphics[width=0.47\textwidth]{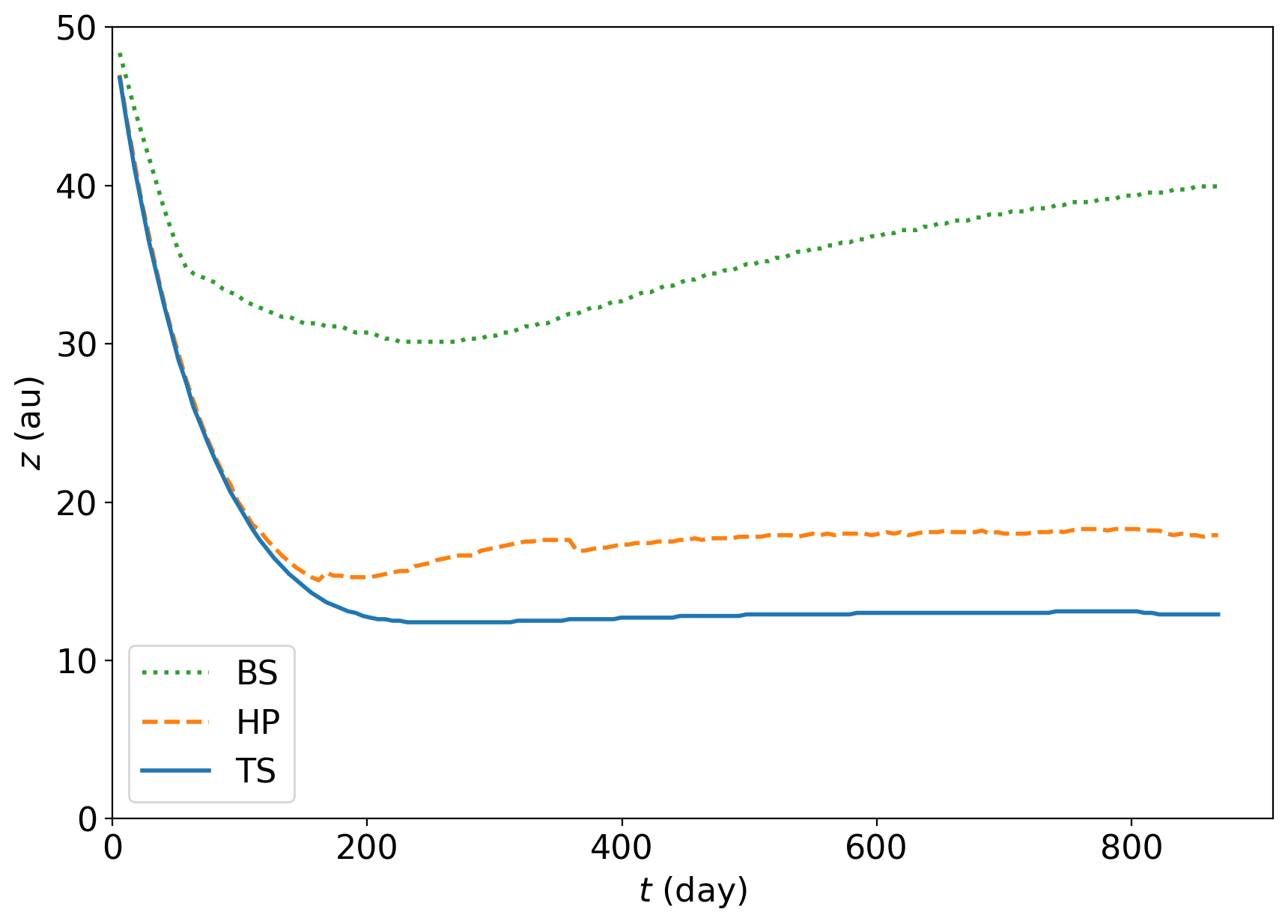}
    \caption{Location of upwind heliosphere features over the full $\sim$900 day simulation for model 6. The TS, HP, and BS are given by the solid blue, dashed orange, and dotted green lines, respectively.}
    \label{fig:shocks_over_time}
\end{figure}

We can apply this Mach number analysis over the duration of the simulation to examine the evolution of discontinuities over time. Figure \ref{fig:shocks_over_time} shows the locations of the TS, HP, and BS for model 6.
Upon reaching the closest approach at $t\sim 250$ days, the TS remains extremely stable.
The HP has some small motion, mostly due to the effect of Kelvin-Helmholtz instabilities.
The BS approaches an innermost position, then slowly retreats upstream over the duration of the simulation. This is a numerical artifact, largely due to an unintended interaction of the BS with the outer $r$-axis. By widening the boundary to capture the full extent of the shock so that the BS falls off the $+r$-direction, we find that the BS retreat is not as dramatic. However, doing so is both computationally more intensive and does not affect the other heliosphere features.

\subsection{Downstream features}
When the TS meets itself at a point in the downwind side of the simulation, it splits into three features labelled in \citet{baranov_model_1993}. These are (from the inside outwards) the Mach disk, the secondary tangential discontinuity, and the reflected shock (as shown in Figure \ref{fig:sims}. In model 1, we even see the appearance of Kelvin-Helmholtz instabilities in this secondary tangential discontinuity. The appearance of these features is a qualitative way to verify the accuracy of this code, as they are expected in the case of a fully-ionized ISM. More detailed models that include a neutral ISM component do not see evidence of these features.

\subsection{Local Bubble density} \label{sec:LB density}
For our Sedov supernova blast, we use a uniform ISM density of $n=0.005\ {\rm cm^{-3}}$. Of course, the Local Bubble does not have that same density everywhere, as evidenced by the Complex of Local Interstellar Clouds. Due to self-similarity in the Sedov blast, the ram pressure does not depend on the ambient density. Therefore, the distance of closest approach should not depend on the ambient density. This relation is also reflected in eq.~(\ref{eqn:rstag_num}), in which ambient density does not appear in the expression for stagnation distance. To verify this relation and examine any other differences beyond stagnation distance, we ran three simulations changing only the ISM ambient density in models 5a-c. The ambient densities used here are $3.34\times 10^{-26}, 1.34\times 10^{-25}$ and $6.68\times 10^{-25}$ g cm$^{-3}$ (i.e., $\rho_0$, $4\rho_0$, and $16\rho_0$) for models 5a, 5b, and 5c, respectively.

\begin{figure*}[!htb]
    \centering
    \begin{interactive}{animation}{animations/SWxSN_63pc_E.mp4}
    \gridline{
        \fig{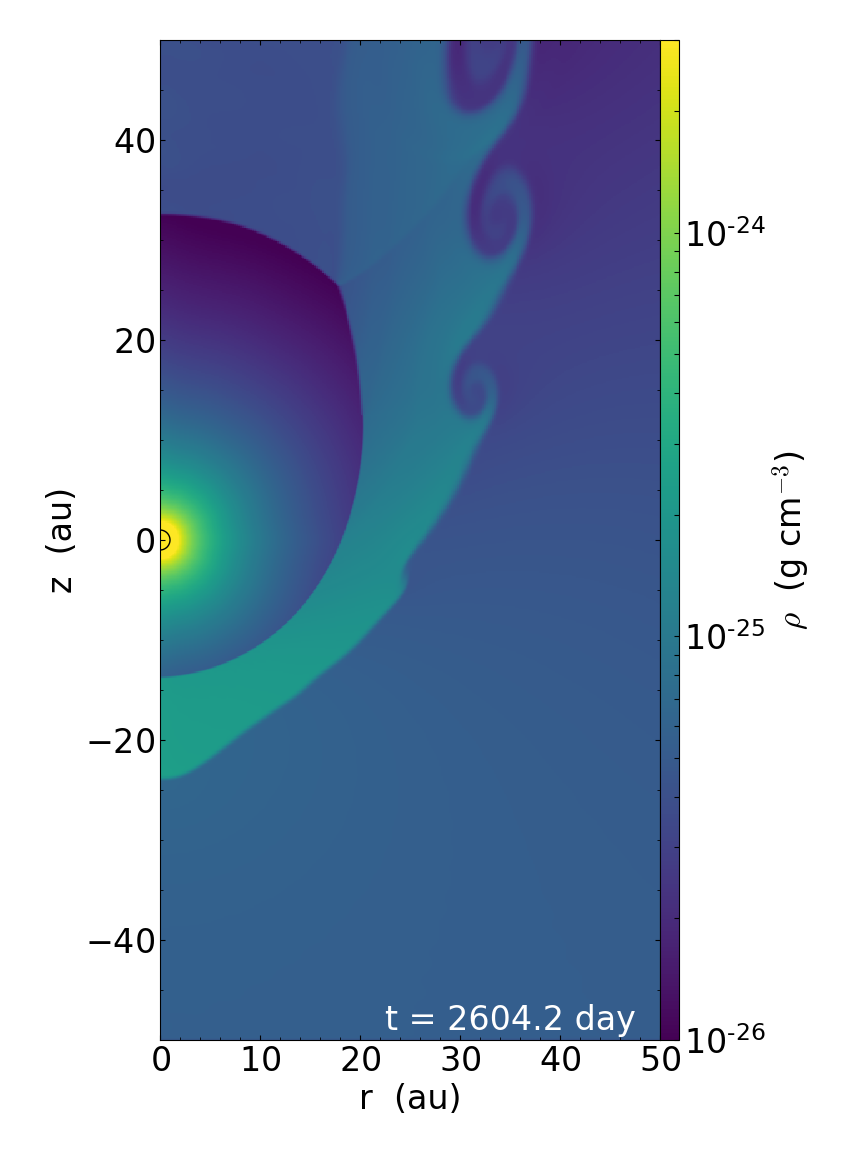}{0.32\textwidth}{(a)}
        \fig{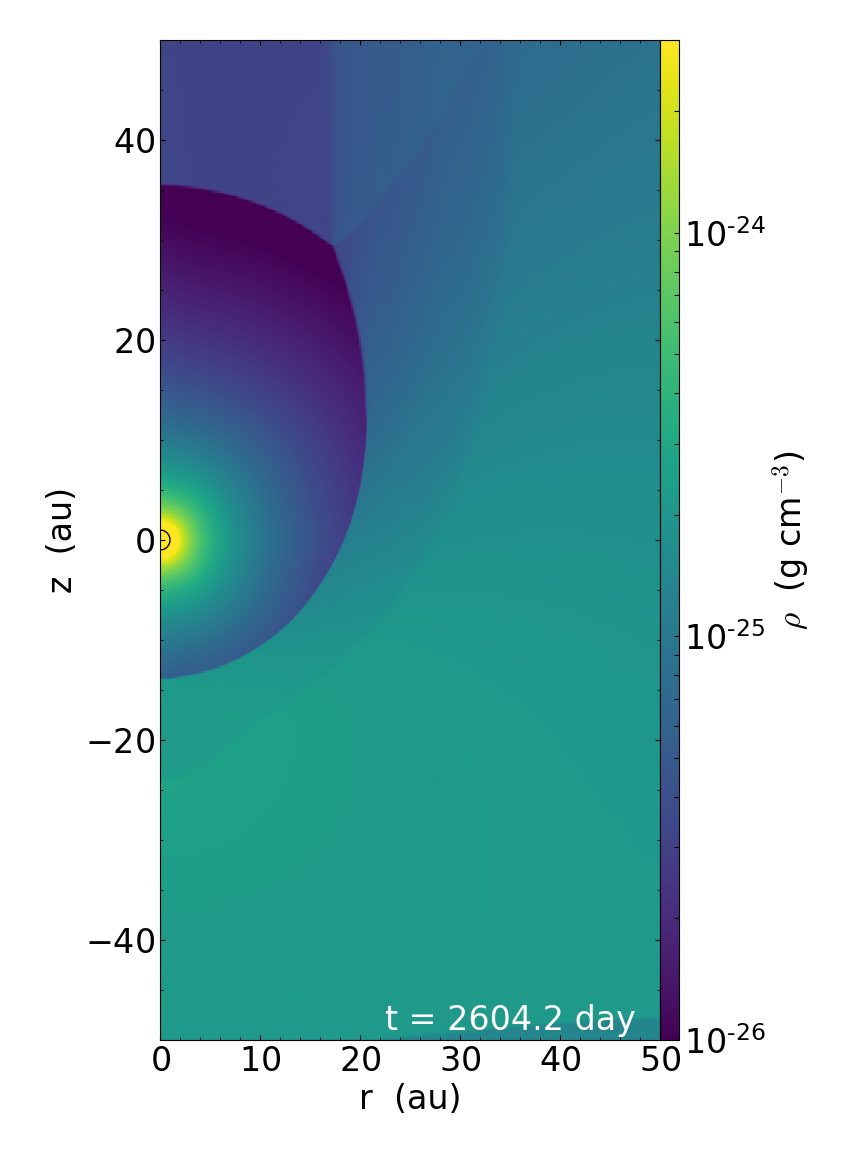}{0.32\textwidth}{(b)}
        \fig{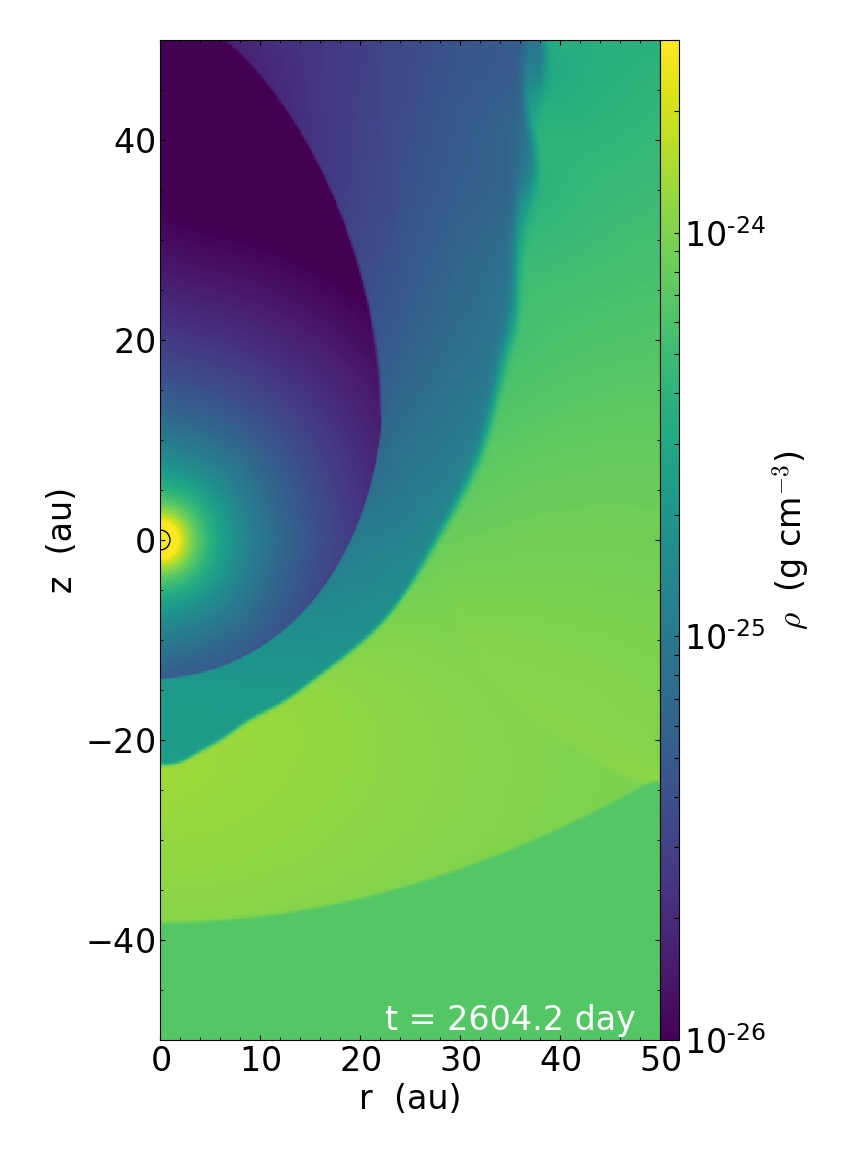}{0.32\textwidth}{(c)}
        }
    \end{interactive}
    \caption{Density plots of models 5a, 5b, and 5c for supernovae 63.3 pc at the same time for three different ambient ISM densities: $3.34\times 10^{-26}, 1.34\times 10^{-25}$ and $6.68\times 10^{-25}$ g cm$^{-3}$, from left to right. An animated version of (a) is available online showing the arrival of the blast wave into the solar system at time $t=0$ and the subsequent compression of the heliosphere until the time of the displayed frame.}
    \label{fig:dens_comparison}
\end{figure*}

We show these simulations after the same amount of time has passed since the introduction of the blast in Figure \ref{fig:dens_comparison} (all are in the isotropic equatorial orientation). The most significant output of these simulations, the distance of closest approach, remains unchanged independent of the ambient ISM density, as expected. Kelvin-Helmholtz instabilities are seen in the low density simulation, but are not as apparent in the higher density cases.
Larger ambient densities slow the blast, resulting in a smaller velocity difference that hinders the growth of these instabilities.
Other features are largely the same, though the smaller blast wave velocity means evolution takes place on a longer timescale. Qualitatively, in Fig. \ref{fig:dens_comparison}(c), we see a clear bow shock and lack of features in the downstream termination shock, but these will continue evolving as the simulation advances.

We conclude that the ambient density does not have a large impact on our simulations. The greatest effect is on the timescale of the heliosphere compression, with a larger ambient density slowing the compression. Larger ISM densities will also increase the time for the blast wave to reach the solar system after the inciting supernova explosion.

\subsection{Orientation effects}

These simulations are performed in two orientations depending on the location of the supernova compared to the equatorial solar wind. The polar wind has a lower density but higher velocity, making the ram pressures similar (see Table \ref{tab:sw-input}). The overall structure of the heliosphere is the same for both orientations. The main difference is that, in the polar orientations, the equatorial wind is bent back through the heliosheath. It does not appear to be a source of instabilities, but reacts to those produced along the HP, as seen in Figure \ref{fig:sims}(b). 

We ran three polar orientations with equatorial counterparts (numbers 3, 6, and 10; model 8 is polar but does not have a corresponding equatorial simulation at the same distance). As expected from the weaker ram pressure, the polar orientation compresses the heliosphere more. This $\sim$10\% difference in the TS, however, is dwarfed by the different values for the supernova distance. We conclude that the supernova distance is more important than its orientation for heliosphere compression.

\section{Discussion and Analysis} \label{sec:discussion}

These simulations of a supernova blast wave colliding with the heliosphere assume idealized hydrodynamics. Although the real situation is surely more complex, we expect that the gross features of the perturbed heliosphere are captured in our simulation.  The apparent lack of a large neutral
component to supernova blasts implies that our single-fluid treatment is a reasonable approximation.
Thus, scaling laws and analytical arguments can be formed.

If one assumes a thin region for the heliosheath, the position of the bow shock should follow a simple analytical expression as a function of angle. As derived by \citet{wilkin_exact_1996}, the position is
\begin{equation}
    r(\theta) = r_{\rm stag} {\rm csc} \theta \sqrt{3(1-\theta {\rm cot} \theta)}.
\end{equation}
Our simulations do not show a thin heliosheath, instead showing a well-separated TS, HP, and BS. Accordingly, this equation is only accurate for small values of $\theta$ directly upstream.

\subsection{Discontinuity Scaling}

Given the final locations of the TS, HP, and BS in each of these simulations, relations between these values and the supernova blast waves can be examined. This comparison will test the usefulness of $r_{\rm stag}$ scaling.

\begin{figure*}[!htb]
    \centering
    \gridline{\fig{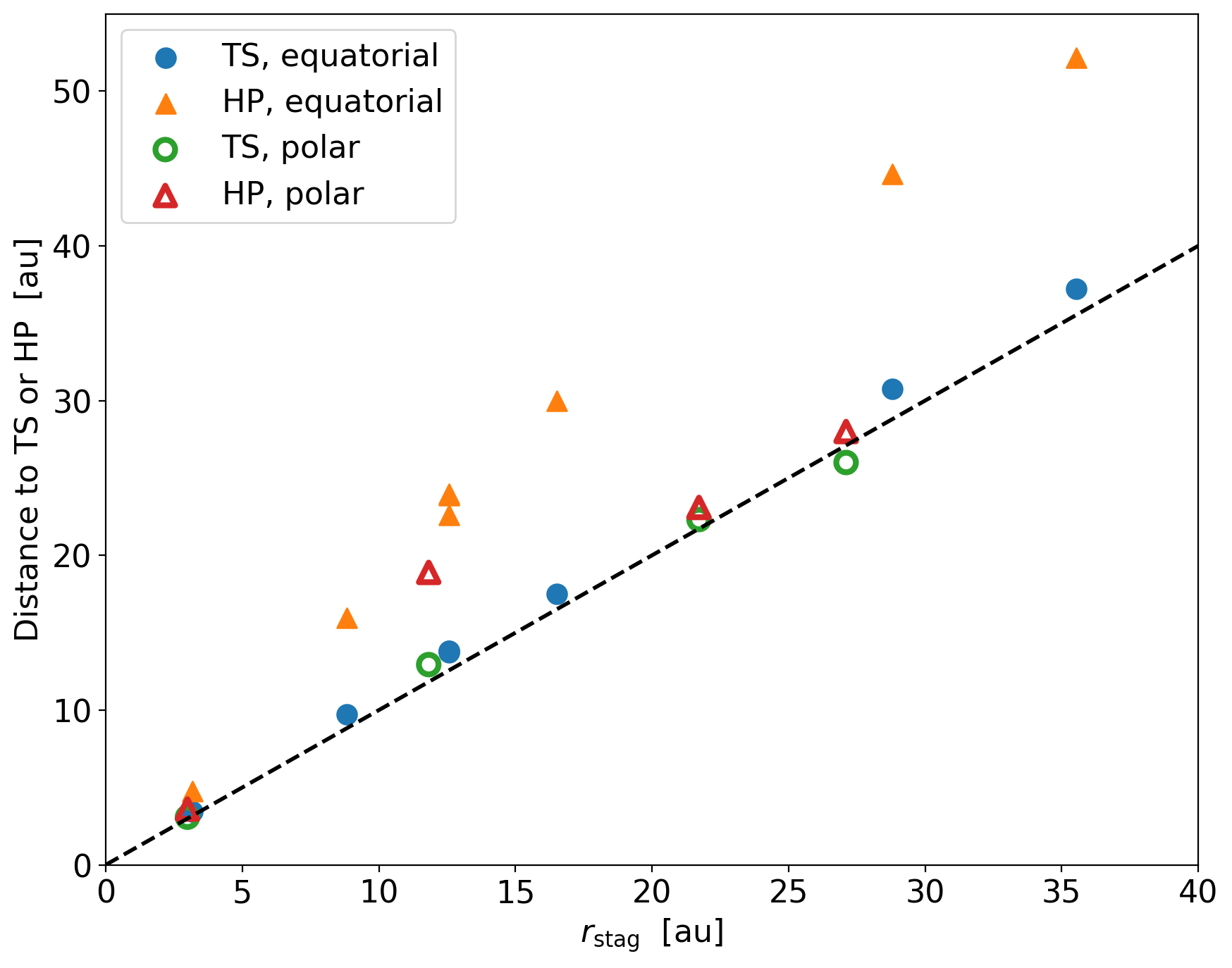}{0.48\textwidth}{(a)}
              \fig{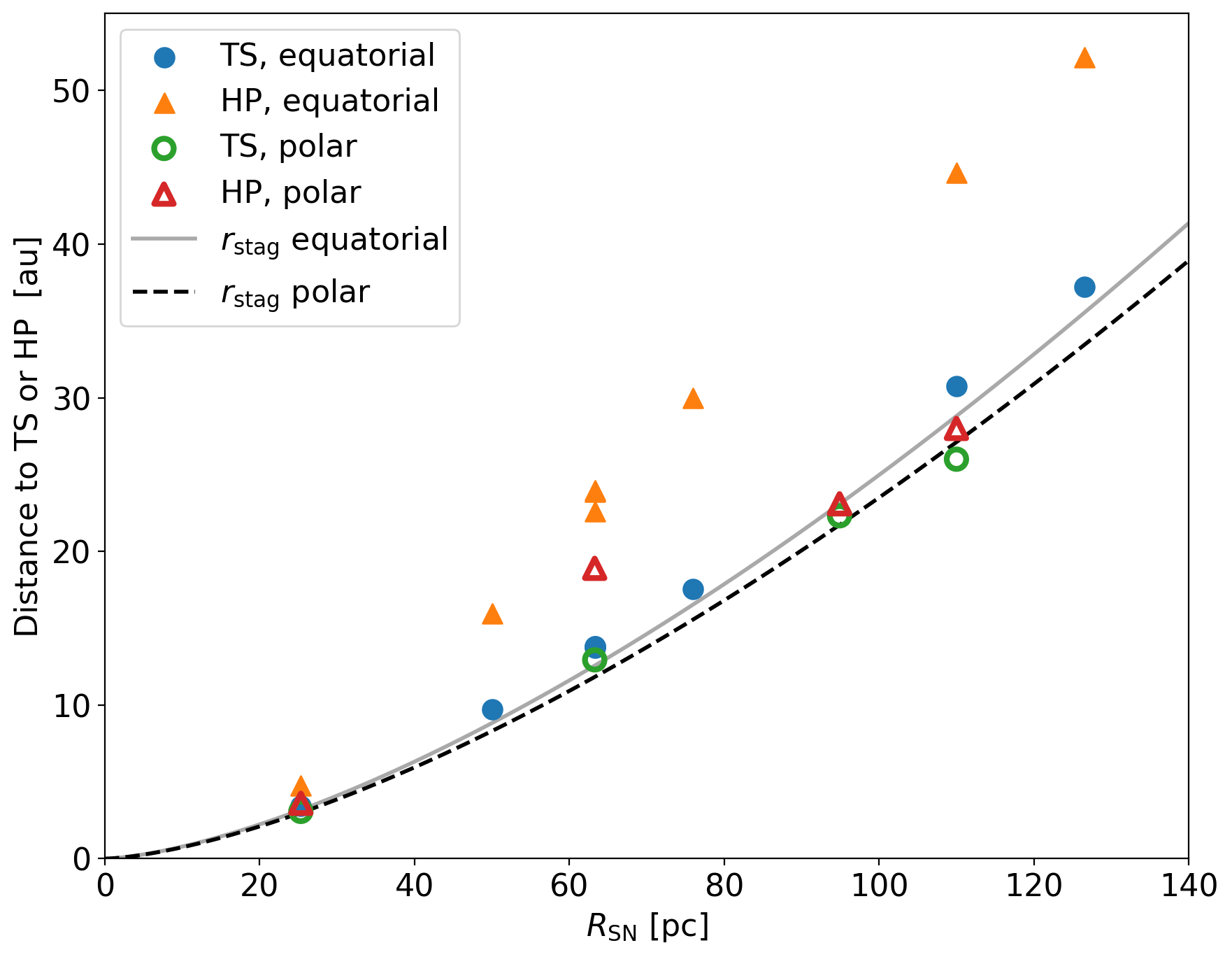}{0.48\textwidth}{(b)}
             }
    \caption{(a) Termination shock and heliopause locations vs stagnation distance from equation \ref{eqn:rstag_num}. The solid blue circles and orange triangles show the TS and HP (respectively) for simulations in the equatorial orientation. The empty green circles and red triangles show the TS and HP from the polar orientation. The dashed black line is a $y=x$ line for reference. (b) The same as (a), but vs $R_{\rm SN}$. The solid grey and dashed black lines are the stagnation distances from Eqn. \ref{eqn:rstag_num} for both equatorial and polar orientations.}
    \label{fig:shock_vs_dists}
\end{figure*}

In Figure \ref{fig:shock_vs_dists}, we compare the location of the TS and HP to two parameters: the stagnation distance from Equation \ref{eqn:r_stag} in (a) and the supernova distance in (b). When comparing to $r_{\rm stag}$, the TS has a very tight relation along the $y=x$ line. According to this relation, $r_{\rm stag}$ is an excellent predictor of the location of the TS rather than the HP. This effect is also noted in, e.g., \citet{comeron_numerical_1998} and \citet{comeron_very_2007}. The location of the HP appears to be dependent on orientation, approaching much closer in the polar orientations (particularly models 8 and 10) than in the equatorial one. We note that these two models also suffered from Kelvin-Helmholtz instabilities near the axis of symmetry. It is possible that the increased grid resolution of these models is responsible for this difference, though it does not affect the models in the equatorial orientation similarly.

Similarly to Figure \ref{fig:shock_vs_dists}(a), we can plot the TS and HP distances as a function of $R_{\rm SN}$, shown in Figure \ref{fig:shock_vs_dists}(b). We also plot eq.~(\ref{eqn:rstag_num}) for both equatorial and polar orientations, though the small difference between these two lines emphasizes how solar orientation is a less significant parameter than supernova distance. We again see the tight correlation between the TS and $r_{\rm stag}$, as well as the very apparent $r_{\rm stag} \propto R_{\rm SN}^{3/2}$ relation.

\begin{figure}[!htb]
    \centering
    \includegraphics[width=0.47\textwidth]{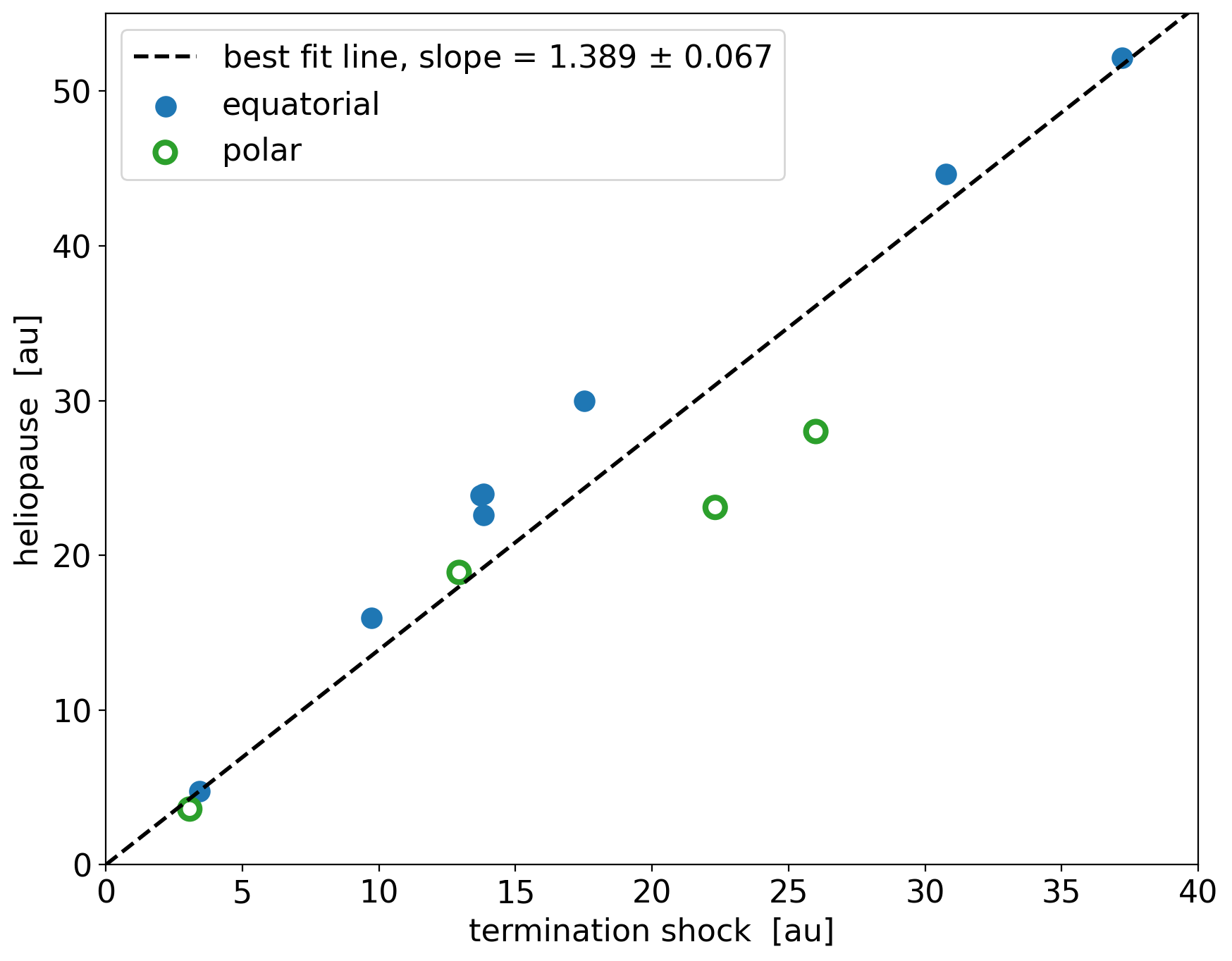}
    \caption{Distance to the heliopause vs termination shock. The simulations in the equatorial orientation are shown with a filled blue circle; the ones in the polar orientation are shown with a empty green circle. The dashed line is the best-fit line to all points, forced to go through the origin.}
    \label{fig:HP_vs_TS}
\end{figure}

The locations of the HP shown in Figure \ref{fig:shock_vs_dists}(a) seem to suggest a correlation between TS and HP distances. To examine this, we plot these two distances against each other in Figure \ref{fig:HP_vs_TS}.
We fit a line to all these points, forcing it to go through the origin. The slope of the best-fit line is $1.389 \pm 0.067$, in very good agreement with \citet{fields_supernova_2008} who found a slope of 1.41.
The smaller slope we find is primarily due to models 8 and 10, which lie well below the line.
As discussed above, this departure from the trend is likely due to the increased resolution, allowing instabilities to form close to the axis of symmetry.
The other models show decent agreement with the fit, though more points would show how well this fit holds over a larger range.

This relation does not hold for the present-day heliosphere, where the TS and HP are $\sim$100 and $\sim$120 au, respectively. The derived relation comes from purely fluid dynamics, and the present-day heliosphere requires a variety of more complex physical process to model correctly. Therefore, the present heliosphere is not required or even expected to fit among this trend.

\subsection{Blast weakening over time}
\label{sect:weaken}

The strength of the supernova blast will weaken over time as the remnant expands. The duration
of this process depends on the explosion distance and local density,
but certainly takes $\gg 1$ kyr.  This timescale is far longer than the $\sim 1$ yr duration of our simulations.
Accordingly, supernova blast properties change on similarly long timescales.
We can study the effect of the weakening supernova blast by considering
``snapshots'' of the blast properties at different times.

We have shown that $r_{\rm stag}$ is an excellent predictor for the distance to the TS.
By taking advantage of this relation, we can extend this analysis over the whole passage of the blast rather than just the leading edge.

The Sedov-Taylor model for supernova remnant evolution can be used to calculate the pressure balance distance for longer time scales. Normally, the Sedov solution is solved in terms of the outermost blast wave. In this instance, we wish to solve for the gas parameters for a stationary observer at a constant location from the origin.

\begin{figure}[!htb]
    \centering
    \includegraphics[width=0.47\textwidth]{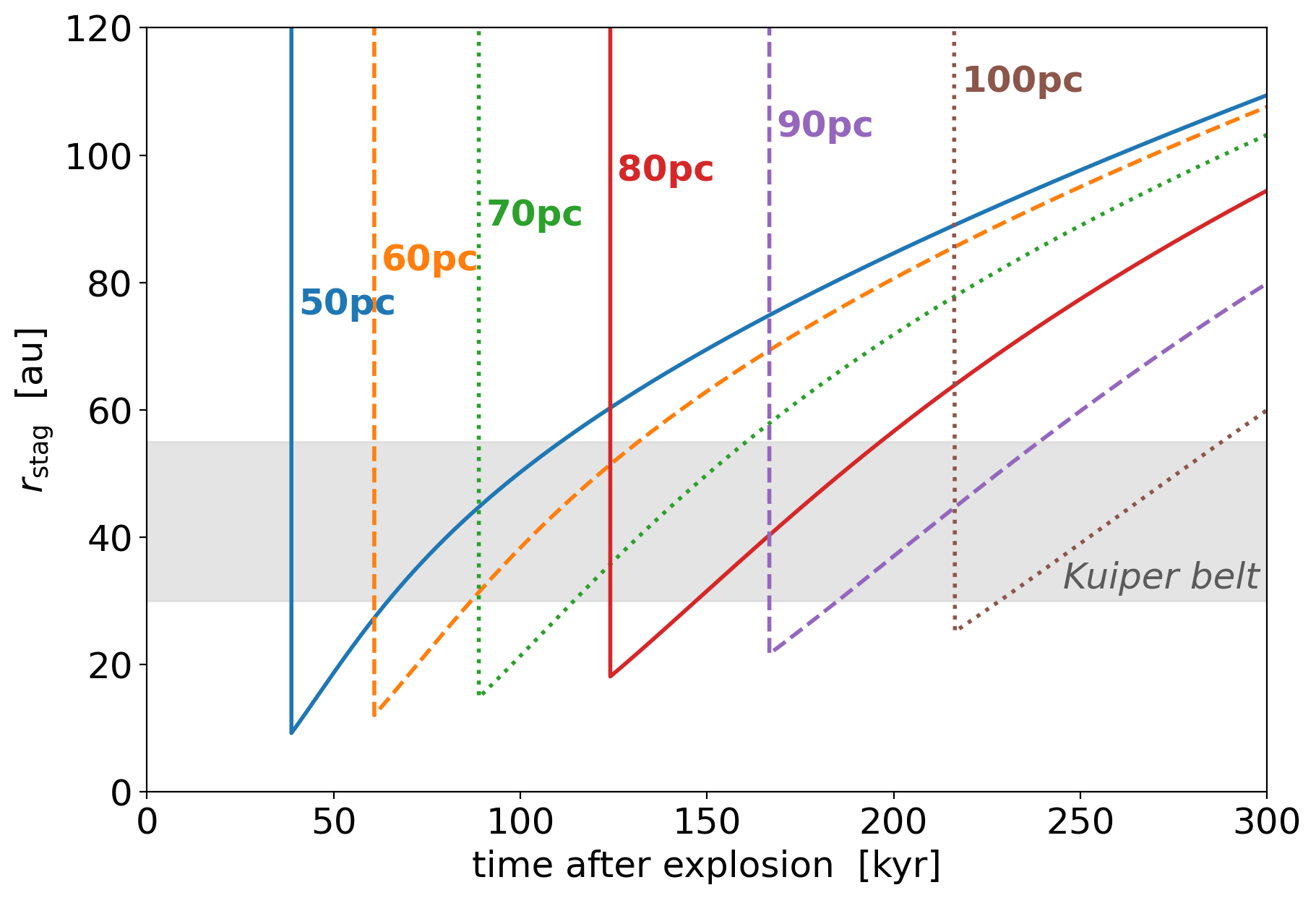}
    \caption{Location of pressure balance for several supernovae at 50-100 pc as the supernova remnant evolves, assuming a Sedov phase and an ISM density of $n_{\rm amb} = 0.005\ {\rm cm}^{-3}$. Vertical line corresponds to the arrival of the forward shock and indicates the
    blast arrival delay after the explosion.  The rebound of the heliosphere thereafter follows the drop in blast ram pressure behind the forward shock.  The shaded band shows the extent of the Kuiper belt. }
    \label{fig:Rstag_over_time}
\end{figure}

We employ a Sedov blast wave verification code\footnote{originally written by Frank Timmes, ported to Python by J. Moskal and J. Workman}
to calculate the Sedov profile over the first 300 kyr after the blast. The chosen ambient density is still that of the Local Bubble,
$n_{\rm amb} = 0.005\ {\rm cm}^{-3}$.
At each timestep, we obtain the thermal and ram pressure and use Eqn. \ref{eqn:r_stag} to calculate $r_{\rm stag}$ (indicating the TS position). 
We show the results of these calculations in Fig. \ref{fig:Rstag_over_time}.
The vertical lines correspond to the initial arrival of the blast; their spacing in time reflects the duration of the blast travel to the Solar System for different supernova distances (eq.~\ref{eq:tarrive}).
Note that on the timescales plotted, our hydrodynamic simulations cover a thin $\sim 1$ yr at the innermost region upon the blast arrival; the rest of the curves use pressure balance to find the closest approach of the supernova. 
Maximum heliospheric compression lasts on the order of $\sim$kyr, however, it takes $>100$ kyr for the blast to weaken enough so that the heliosphere fully rebounds.

Figure~\ref{fig:Rstag_over_time} has important implications for delivery of \fe60 and other radioisotopes to the Earth and Moon.  We see that the closest approach of the blast recedes quite rapidly at first.  If the supernova ejecta is well-mixed into and carried by the blast, this means that the distance it must travel through the heliosphere rapidly becomes larger over time.  If instead the ejecta is not well-mixed but located well behind the forward shock, it has even farther to travel in the heliosphere than in the well-mixed case.  This again points to the need for the radioisotopes to arrive on dust grains having large sizes or high speeds.

After the forward shock arrival, much of the solar system will be exposed to the blast wave. Fig. \ref{fig:Rstag_over_time} also shows the region of the Kuiper belt from 30 to 55 au \citep{sanctis_thermal_2001, chiang_resonance_2003}. Even for the most distant 100 pc supernova, the entirety of the Kuiper belt is exposed for $\sim$10 kyr.

It is interesting to note that due to the slightly sharper weakening of the closer supernovae, the outer Kuiper belt at 55 au is exposed to the supernova blast for approximately 70 kyr regardless of distance. In contrast, the inner Kuiper belt at 30 au is more sensitive to the distance, resulting in a nearly linear relation with distance.

\subsection{Other solar system effects}

Given the extent to which the heliosphere can be compressed, parts of the outer solar system are directly exposed to the supernova blast. This exposure may have numerous effects on the outer bodies. \citet{stern_influence_1988} investigated how the light from a nearby supernova could melt the outermost surface of comets. \citet{stern_ism-induced_1990} calculated the erosion of these bodies due to SNRs and how small particles with radius $\lesssim$ 100 $\mu$m would be ejected from the solar system. Both of these studies can now be re-contextualized in light of the known supernovae detected by terrestrial \fe60: the most recent 3 Myr supernova may have ``cleaned'' the Oort cloud of small dust grains but left larger comet orbits unperturbed.

Due to the increased abundance of cosmic rays, the cosmic ray exposure ages of asteroid surfaces may be affected. While typical isotopes of interested for exposure ages have half-lives of less than $\lesssim$ 1 Myr \citep[see, e.g.,][]{michel_cross_1997}, some of the longer-lived radioisotopes like \fe60, \mn53, and \al26 could have high abundances. This effect would be more apparent in metallic meteoroids, which are exposed for longer than stony meteoroids \citep{ammon_new_2009}.

Potential links between SNRs and planets have so far been almost wholly unexplored. A significant challenge is to suggest not only how a supernova affects planets, but also find what observable features could remain after millions of years. We leave these investigations to future studies.

\subsection{Supernova effects on astrospheres}

In these simulations, we examine the supernova blast wave's effect on our own heliosphere. These simulations can be generalized to examine the effect of supernovae on astrospheres, which are driven by stellar winds from other stars.

Astrospheres are commonly observed by their bow shocks in the IR part of the spectrum. They are most commonly seen in surveys, either by the rapidly-moving runaway stars \citep{peri_e-boss_2012} or in the dusty environment of the Galactic plane \citep{kobulnicky_comprehensive_2016}.
Red supergiants in particular, such as Betelgeuse, are expected to have enormous astrospheres nearly a parsec wide \citep{meyer_3d_2021}.
As of yet, no bow shocks have been associated with a star located within an SNR.
Indeed, supernovae are prolific destroyers of ISM dust, so it is perhaps expected that astrospheres in SNRs would not be detectable in the IR due to a lack of dust.
However, if discovered, they would be a novel form of stellar-interstellar interaction.

Astrospheres present an interesting complementary aspect to our own heliosphere: though we can directly probe our own heliosphere, the overall shape of the heliopause is still debated. It may have a comet-like tail 1000s of au long \citep[e.g.,][]{izmodenov_three-dimensional_2015} or it may be truncated much closer \citep{opher_magnetized_2015}. In contrast, known astrospheres cannot be probed directly, but their shape can be seen by their bow shocks.

Both the solar system penetration distance and the increase in cosmic rays may alter the potential habitability of astrospheres. The proximity of stellar systems to supernovae affects the Galactic Habitable Zone \citep{gonzalez_galactic_2001, lineweaver_galactic_2004, morrison_extending_2015,spinelli_habitable_2021}. With these simulations, more accurate calculations of cosmic ray exposure during supernova blasts can be made in order to better ascertain astrosphere viability.

\subsection{Effects of solar motion}

We have assumed the solar system is at rest relative to the supernova explosion.  In general one expects the Sun will move relative to the supernova progenitor and blast center.  A nonzero solar velocity relative to the blast would change the ram pressure seen by the heliosphere, and the impact of this change scales at $\delta P_{\rm ram}/P_{\rm ram} \sim v_\odot/v_{\rm blast}$.  The Sun's present motion with respect to the stellar Local Standard of Rest is 18 km/s \citep{schonrich_local_2010, zbinden_simple_2019}
and our speed relative to the very local ISM is 27 km/s; for such values $v_\odot/v_{\rm blast} \ll 1$ when blast arrives, and the perturbation is small.  

At late times, the blast speed slows and Earth's speed could become important; these effects are discussed in \citep{chaikin_simulations_2021} in the context of \fe60 deposition and the local environment encountered by the solar system.  It remains for future work to model such effects on the heliosphere, including the possibility that the Earth's velocity is misaligned with that of the supernova blast.

\section{Conclusions} \label{sec:conclusion}

Motivated by terrestrial detections of \fe60 as evidence for near-Earth supernovae in the recent past, we have presented hydrodynamic simulations of the heliosphere's response to a supernova blast wave at various distances. We match our steady solar wind to that observed by space missions and test the effect of solar wind orientation. The supernova blast is assumed to be in the Sedov phase.

The broad structure of the heliosphere is reproduced, albeit at much smaller scales than the present-day heliosphere. We verified that pressure balance gives the location of the TS over a range of distances. We applied this relation to analytically examine the heliosphere throughout the duration of the supernova remnant evolution far longer than the hydrodynamic simulations could run.

Taking advantage of rotational axisymmetry allowed for two orientations of how the blast wave strikes the heliosphere:
polar, in which the blast approaches from the poles of the Sun, and equatorial, in which the blast arrives from the side.
We apply a steady fast and slow solar wind originating from the poles and equator, respectively.
Due to their density differences, the ram pressures of these winds are very similar.
Appropriately, since the penetration distance depends chiefly on ram pressure, the orientation is found to have little influence on the global structure of the heliosphere.

This work reaffirms and builds upon the conclusions of \citet{fields_supernova_2008}
that the supernova blast plasma is strongly excluded from 1 au for any plausible distance to the
recent \fe60-depositing supernovae.  The observed
\fe60 deposits on the Earth and Moon
must have arrived in a form other than the plasma--namely, in dust grains.
The dynamics of dust grains in the outer heliosphere are well-studied \citep[e.g.,][]{belyaev_dynamics_2010}, especially for the present-day heliosphere.
\citet{wallis_penetration_1987} found that during the passage through a dense cloud, dust can penetrate the heliosphere to Earth with little deflection due to the heliosphere's small size.
\citet{athanassiadou_penetration_2011, fry_radioactive_2016} found that dust grains from near-Earth supernovae are typically deflected less than 1$^\circ$ by the heliosphere. 
Our work shows provides the location of closest approach of the supernova material
and thus the initial conditions for studies of the dust propagation within the compressed heliosphere to
Earth.  

As the SNR evolves, the blast wave will weaken and allow the heliosphere to rebound.
According to our scaling laws, this process is expected to take several $100\ \rm kyr$ to rebound to 100 au.
Our simulations do not account for supernova-formed dust dynamics, but their propagation through the heliosphere is modified by the decreased heliosphere size.
This effect is especially significant for any grains that arrive within the first 100 kyr after the blast wave arrival.  Grains that arrive later will need to traverse progressively more of the heliosphere in order to reach the Earth and Moon.

The principal shocks in our simulations (BS and TS) can accelerate particles through diffusive Fermi acceleration to produce anomalous cosmic rays \citep{zank_interstellar_1996, lazarian_model_2009}.
Cosmic rays accelerated in these shocks would have an effect on the amount of radiation impinging on Earth. For very nearby supernovae, there could even be biological effects, either as a mass extinction \citep{gehrels_ozone_2003} or lesser extinction events \citep{melott_supernova_2017, thomas_terrestrial_2016}.
Tracking how these shocks evolve over the passage of the blast wave furthers our understanding of potential biological responses to the supernova.

The proposed Interstellar Probe mission\footnote{\url{https://interstellarprobe.jhuapl.edu/}} plans to launch a spacecraft several hundred au into the very local ISM in the coming decades. Such a probe would be sent out to the remains of an ancient supernova remnant, and may thus contribute to the study of how old SNRs evolve and fade into the Galactic medium.

The simulations presented here could be expanded upon in many ways, including developing a more careful treatment of non-hydrodynamic physics like magnetic fields and charge exchange. Solar activity represented as a time-varying solar wind may also be relevant for determining how instabilities affect the distance of closest approach. Such inclusions would allow for more realistic non-axisymmetric simulations, allowing us to probe how our own solar system responds to dramatic nearby events such as supernovae.

\acknowledgments
The work of J.A.M. was supported by the Future Investigators in NASA Earth and Space Science and Technology (FINESST) program under award number 80NSSC20K1515. The work of B.D.F. was supported in part by the NSF under grant number AST-2108589.
We gratefully acknowledge helpful discussions with: John Ellis, Adrienne Ertel, Brian Fry, Zhenghai Liu, Phil Coady, and Leeanne Smith about near-Earth supernovae; Pontus Brandt, Merav Opher, and Elena Provornikova about heliosphere-ISM interactions. We thank the creators and contributors of the \texttt{Athena++} code, particularly Patrick Mullen for his direct assistance.

We acknowledge the use of the NASA National Space Science Data Center and the Space Physics Data Facility OMNIWeb Database for the following: {\em Voyager 2} PLS data (PI John W. Belcher), and {\em Ulysses} SWOOPS data (PI Dave J. McComas).

\software{\\
    Athena++ \citep{stone_athena_2020},\\
    yt \citep{turk_yt_2011},\\
    Numpy \citep{harris_array_2020},\\
    Matplotlib \citep{hunter_matplotlib_2007},\\
    Sedov verification (originally written by F. Timmes, ported to Python by J. Moskal and J. Workman and available at: \url{https://org.coloradomesa.edu/~jworkman/})
}

\bibliography{biblib.bib}
\bibliographystyle{aasjournal}

\end{document}